%% file: ms.tex
\documentclass[lettersize]{emulateapj}

\input{lyaf_macros}

\slugcomment{To be submitted to AJ}
\shorttitle{BOSS DR9 \lya\ Forest Sample}
\shortauthors{Lee \etal}

\begin{document}

\title{The BOSS Lyman-$\alpha$ Forest Sample from SDSS Data Release 9}
\include{authors}

\begin{abstract}
We present the BOSS Lyman-$\alpha$ (\lya) Forest Sample from SDSS Data Release 9, comprising 54,468
quasar spectra with $\zq > 2.15$ suitable for \lya\ forest analysis. This data set probes the intergalactic medium with absorption redshifts
 $2.0 < \zalp < 5.7$
over an area of 3275 square degrees, and encompasses an approximate comoving volume of $20\;h^{-3}\,{\rm Gpc^3}$.  
With each spectrum, we have included several products 
designed to aid in \lya\ forest analysis: 
improved sky masks that flag pixels where data may be unreliable, corrections for known biases
in the pipeline estimated noise, masks for the cores of damped \lya\ systems and corrections
for their wings, and estimates of the unabsorbed continua so that the observed flux can be converted
to a fractional transmission.
The continua are derived using a principal component fit to the
quasar spectrum redwards of restframe \lya\ ($\lambda > 1216\;\ang$),
extrapolated into the forest region and normalized by a linear function to fit the 
expected evolution of the 
\lya\ forest mean-flux.  The estimated continuum errors are $\lesssim 5 \%$ rms.
We also discuss possible systematics arising from uncertain spectrophotometry and 
artifacts in the flux calibration; global corrections for the latter are provided. 
Our sample provides
a convenient starting point for users to analyze clustering
in BOSS \lya\ forest data, and it provides a fiducial data set that
can be used to compare results from different analyses of baryon
acoustic oscillations in the \lya\ forest.
The full data set is available from the SDSS-III DR9
web site.
\end{abstract}

\keywords{intergalactic medium --- quasars: emission lines --- 
quasars: absorption lines --- methods: data analysis}

\input{intro}

\input{boss_summary}

\input{selection}

\input{value_added}

\input{systematics}

\input{conclusions}

\acknowledgements{
Funding for SDSS-III has been provided by the Alfred P. Sloan Foundation, the Participating Institutions, the National Science Foundation, and the U.S. Department of Energy Office of Science. The SDSS-III web site is \url{http://www.sdss3.org/}.

SDSS-III is managed by the Astrophysical Research Consortium for the Participating Institutions of the SDSS-III Collaboration including the University of Arizona, the Brazilian Participation Group, Brookhaven National Laboratory, University of Cambridge, Carnegie Mellon University, University of Florida, the French Participation Group, the German Participation Group, Harvard University, the Instituto de Astrofisica de Canarias, the Michigan State/Notre Dame/JINA Participation Group, Johns Hopkins University, Lawrence Berkeley National Laboratory, Max Planck Institute for Astrophysics, Max Planck Institute for Extraterrestrial Physics, New Mexico State University, New York University, Ohio State University, Pennsylvania State University, University of Portsmouth, Princeton University, the Spanish Participation Group, University of Tokyo, University of Utah, Vanderbilt University, University~of~Virginia, University~of~Washington, and Yale~University.   \
 }
\vspace{1em}
% List of sky lines and metal ions in Appendix?

\bibliographystyle{apj}
\bibliography{lyaf_kg,apj-jour}
\end{document}

%% file: lyaf_macros.tex
\usepackage{natbib}
\usepackage{amsmath}
\usepackage{CJK}
\usepackage{epsf}

\newcommand{\alphat}{\ensuremath{\alpha}}

\newcommand{\mpc}{\ensuremath{\, h^{-1}\,\mathrm{Mpc} }}

\newcommand{\kms}{\, \mathrm{km \; s^{-1}}}
\newcommand{\snr}{\ensuremath{\mathrm{S/N}}}
\newcommand{\lya}{Ly$\alpha$}
\newcommand{\lyb}{Ly$\beta$}
\newcommand{\waveion}[3]{\ion{#1}{#2} $\lambda$#3}
\newcommand{\wavelya}{Ly$\alpha \; \lambda 1216$}

\newcommand{\fmean}{\ensuremath{\langle F \rangle}}

\newcommand{\fluxergs}{\mathrm{erg \;s^{-1} cm^{-2} \ang^{-1}}}

\newcommand{\npix}{\ensuremath{N_\mathrm{pix}}}
\newcommand{\lambdamax}{\ensuremath{\lambda_\mathrm{max}}}
\newcommand{\lambdamin}{\ensuremath{\lambda_\mathrm{min}}}

\newcommand{\lambrest}{\ensuremath{\lambda_\mathrm{rest}}}

\newcommand{\zq}{\ensuremath{z_\mathrm{qso}}}
\newcommand{\zabs}{\ensuremath{z_\mathrm{abs}}}

\newcommand{\zalp}{\ensuremath{z_{\alpha}}}

\newcommand{\ang}{\ensuremath{\mathrm{\AA}}}
\newcommand{\cpca}{\ensuremath{C_\mathrm{PCA}}}
\newcommand{\cmf}{\ensuremath{C_\mathrm{MF}}}

\newcommand{\fpipe}{\ensuremath{f_{\rm p}}}
\newcommand{\wpipe}{\ensuremath{w_{\rm p}}}
\newcommand{\sigpipe}{\ensuremath{\sigma_{\rm p}}}

\newcommand{\fpipei}{\ensuremath{f_{{\rm p},i}}}
\newcommand{\wpipei}{\ensuremath{w_{{\rm p},i}}}
\newcommand{\sigpipei}{\ensuremath{\sigma_{{\rm p},i}}}

\newcommand{\wfi}{\ensuremath{w_{{\rm F},i}}}

\newcommand{\beq}{\begin{equation}}
\newcommand{\eeq}{\end{equation}}
\newcommand{\bc}{\begin{center}}
\newcommand{\ec}{\end{center}}
\newcommand{\bfig}{\begin{figure}}
\newcommand{\efig}{\end{figure}}

\def\etal   {et~al.}

%% file: authors.tex
\author{
Khee-Gan~Lee\altaffilmark{1},
Stephen~Bailey\altaffilmark{2},
Leslie~E.~Bartsch\altaffilmark{3,4},
William~Carithers\altaffilmark{2},
Kyle~S.~Dawson\altaffilmark{5},
David~Kirkby\altaffilmark{6},
Britt~Lundgren\altaffilmark{7},
Daniel~Margala\altaffilmark{6},
Nathalie~Palanque-Delabrouille\altaffilmark{8}, 
Matthew~M.\ Pieri\altaffilmark{9},
David J. Schlegel\altaffilmark{2},
David H.\ Weinberg\altaffilmark{10},
Christophe~Y\`eche\altaffilmark{8},
\'{E}ric~Aubourg\altaffilmark{11},
Julian~Bautista\altaffilmark{11},
Dmitry~Bizyaev\altaffilmark{12},
Michael~Blomqvist\altaffilmark{6},
Adam~S.\ Bolton\altaffilmark{5},  
Arnaud~Borde\altaffilmark{6},
Howard~Brewington\altaffilmark{12},
Nicol\'as~G.\ Busca\altaffilmark{11}, 
Rupert~A.C.\ Croft\altaffilmark{3,4},
Timoth\'ee Delubac\altaffilmark{8}, 
Garrett~Ebelke\altaffilmark{12},
Daniel~J.\ Eisenstein\altaffilmark{13},
Andreu~Font-Ribera\altaffilmark{2,14},
Jian~Ge\altaffilmark{15},
Jean-Christophe~Hamilton\altaffilmark{11},
Joseph~F.\ Hennawi\altaffilmark{1},
Shirley~Ho\altaffilmark{3,4},
Klaus~Honscheid\altaffilmark{16},
Jean-Marc~Le~Goff\altaffilmark{8}, 
Elena ~Malanushenko\altaffilmark{12},
Viktor~Malanushenko\altaffilmark{12},
Jordi~Miralda-Escud\'{e}\altaffilmark{17,18},
Adam~D.\ Myers\altaffilmark{19},
Pasquier~Noterdaeme\altaffilmark{20},
Daniel~Oravetz\altaffilmark{12},
Kaike~Pan\altaffilmark{12},
Isabelle~P\^aris\altaffilmark{20},
Patrick~Petitjean\altaffilmark{20},
James~Rich\altaffilmark{8},
Emmanuel~Rollinde\altaffilmark{20},
Nicholas~P.\ Ross\altaffilmark{2},
Graziano~Rossi\altaffilmark{8,9},
Donald~P.\ Schneider\altaffilmark{21,22},
Audrey~Simmons\altaffilmark{12},
Stephanie~Snedden\altaffilmark{12},
An\v{z}e Slosar\altaffilmark{23},
David~N.\ Spergel\altaffilmark{24},
Nao~Suzuki\altaffilmark{2},
Matteo~Viel\altaffilmark{25,26},
Benjamin~A.\ Weaver\altaffilmark{27} 
}

\altaffiltext{1}{Max Planck Institute for Astronomy, K\"{o}nigstuhl 17, 69115 Heidelberg, Germany}
\altaffiltext{2}{Lawrence Berkeley National Lab, 1 Cyclotron Rd, Berkeley, CA 94720, USA}
\altaffiltext{3}{Department of Physics, Carnegie Mellon University, 5000 Forbes Ave, Pittsburgh, PA 15213}
\altaffiltext{4}{Bruce and Astrid McWilliams Center for Cosmology, Carnegie Mellon University, Pittsburgh, PA 15213, USA}
\altaffiltext{5}{Department of Physics and Astronomy, University of Utah, 115 S 1400 E, Salt Lake City, UT 84112, USA}
\altaffiltext{6}{Department of Physics and Astronomy, University of California, Irvine, CA 92697, USA}
\altaffiltext{7}{Department of Astronomy, University of Wisconsin, 475 North Charter Street, Madison, WI 53706, USA}
\altaffiltext{8}{CEA, Centre de Saclay, Irfu/SPP, F-91191 Gif-sur-Yvette, France} 
\altaffiltext{9}{Institute of Cosmology and Gravitation, Dennis Sciama Building, University of Portsmouth, Portsmouth, PO1 3FX, UK}
\altaffiltext{16}{Department of Astronomy and Center for Cosmology and Astro-Particle Physics, Ohio State University, Columbus, OH 43210, USA}
\altaffiltext{11}{APC, Universit\'{e} Paris Diderot-Paris 7, CNRS/IN2P3, CEA, Observatoire de Paris, 10, rueA. Domon \& L. Duquet,  Paris, France}
\altaffiltext{12}{Apache Point Observatory, P.O. Box 59, Sunspot, NM 88349, USA}
\altaffiltext{13}{Harvard-Smithsonian Center for Astrophysics, Harvard University, 60 Garden St., Cambridge MA 02138, USA}
\altaffiltext{14}{Institute of Theoretical Physics, University of Zurich, 8057 Zurich, Switzerland }
\altaffiltext{15}{Department of Astronomy, University of Florida, Bryant Space Science Center, Gainesville, FL 32611-2055, USA}
\altaffiltext{16}{Department of Physics and Center for Cosmology and Astro-Particle Physics, Ohio State University, Columbus, OH 43210, USA}
\altaffiltext{17}{Instituci\'{o} Catalana de Recerca i Estudis  Avan\c{c}ats, Barcelona, Catalonia}
\altaffiltext{18}{Institut de Ci\`{e}ncies del Cosmos, Universitat de Barcelona/IEEC, Barcelona 08028, Catalonia}
\altaffiltext{19}{Department of Physics and Astronomy, University of Wyoming, Laramie, WY 82071, USA}
\altaffiltext{20}{Universit\'e Paris 6 et CNRS, Institut d'Astrophysique de Paris, 98bis blvd. Arago, 75014 Paris, France}
\altaffiltext{21}{Department of Astronomy and Astrophysics, The Pennsylvania State University, University Park, PA 16802}
\altaffiltext{22}{Institute for Gravitation and the Cosmos, The Pennsylvania State University, University Park, PA 16802}
\altaffiltext{23}{Bldg 510 Brookhaven National Laboratory, Upton, NY 11973, USA}
\altaffiltext{24}{Department of Astrophysical Sciences, Princeton University, Princeton, NJ 08544, USA}
\altaffiltext{25}{INAF, Osservatorio Astronomico di Trieste, Via G. B. Tiepolo 11, 34131 Trieste, Italy}
\altaffiltext{26}{INFN/National Institute for Nuclear Physics, Via Valerio 2, I-34127 Trieste, Italy}
\altaffiltext{27}{Center for Cosmology and Particle Physics, New York University, New York, NY 10003 USA}
\email{lee@mpia.de}

%% file: intro.tex
\section{Introduction}

The Lyman-\alphat\ (\lya) forest \citep{lynds:1971} is the ubiquitous absorption pattern observed in the spectra of high-redshift 
quasars, caused by \wavelya\ absorption of residual neutral hydrogen embedded in a highly photo-ionized 
($n_{HI}/n_{H} \lesssim 10^{-5}$) intergalactic medium \citep[see, e.g., ][]{gunn:1965, rauch:1998, meiksin:2009}. 
Over the past two decades, studies using both numerical and semi-analytic methods have established 
that the \lya\ forest directly traces the underlying dark matter fluctuations in inter-galactic space
\citep{cen:1994, bi:1995, zhang:1995, hernquist:1996, miralda-escude:1996, bi:1997, hui:1997, theuns:1998}. 
This theoretical insight has enabled the \lya\ forest to be used as a cosmological probe of the
high-redshift ($z \gtrsim 2$) universe \citep[e.g.,][]{croft:1998, mcdonald:2000, croft:2002, zaldarriaga:2003, viel:2004, mcdonald:2005,mcdonald:2006}.  

In particular, the picture of the \lya\ forest as a continuous tracer of the underlying dark matter density
implies that observers no longer must resolve individual forest lines to measure large-scale 
correlations \citep{croft:1998, weinberg:2003} ---
this advance enables the use of moderate-resolution spectra that do not fully resolve the \lya\ forest absorption to 
perform measurements of large-scale structure at $z \gtrsim 2$. 
\citet{mcdonald:2006} used a sample of 3035 moderate resolution \lya\ forest spectra from the
Sloan Digital Sky Survey \citep[SDSS, ][]{york:2000} to measure the 1-dimensional flux power spectrum at 
$z = 2.2-4.2$, allowing constraints to be placed on the linear matter power spectrum \citep{mcdonald:2005, seljak:2005}
and neutrino masses \citep{seljak:2006}.  
At higher quasar sightline densities, correlations can be measured in the transverse direction across different sightlines.
\citet{mcdonald:2007}  proposed that three-dimensional measurements of the \lya\ forest flux correlation 
could be used to measure the baryon acoustic oscillation (BAO) signature at scales of $\sim 100 \mpc$.

One of the key goals of the Baryon Oscillation Spectroscopic Survey \citep[BOSS,][]{dawson:2012}, 
of SDSS-III \citep{eisenstein:2011} is to carry out  
precision BAO measurements from the \lya\ forest at $z \approx 2.5$; 
for recent cosmological results from the BOSS {\it galaxy} redshift survey see, e.g.,
\citet{anderson:2012, sanchez:2012, reid:2012}.
Over its projected 4.5-year survey period, BOSS aims to obtain spectra of $170,000$ quasars with $z \gtrsim 2$, 
with an areal density of $15-20 \; \mathrm{deg}^{-2}$. The first public release of BOSS spectra was
through SDSS Data Release 9 \citep[DR9,][]{sdss:2012} in July 2012,
comprising the first 1.5 years of BOSS observations spanning Dec 2009 - July 2011.  
DR9 comprises 535,995 new galaxy spectra and 102,100 quasar spectra at all redshifts, covering
$3275 \;\mathrm{deg}^{2}$ of the sky. 
At the time of writing, the BOSS data have already provided the first
measurements of large-scale 3-dimensional correlations in the \lya\ forest \citep{slosar:2011}, 
and we have recently reported the first BAO detection from the \lya\ forest \citep{busca:2012}, 
yielding a measurement of the expansion rate at $t\approx 3\,$Gyr,
intermediate between the recombination epoch probed by the cosmic microwave background
and the ``acceleration era'' beginning at $z \approx 0.8$, or $t \approx 6\,$Gyr.
Because \lya\ forest BAO measurement is a novel endeavor and a central goal of BOSS,
the collaboration is carrying out the first analyses using two largely independent
methodologies and codes; results from the alternative BAO analysis will be presented
by \citet{slosar:2012}.

The spectra used in these papers are all available via DR9, and the
DR9 quasar catalog is described and presented by \cite{paris:2012}.
However, there are a number of complex steps between a set of quasar
spectra and a cosmological analysis of the \lya\ forest, including flagging
unreliable data, removing or correcting regions affected by damped \lya\ 
absorbers (DLAs) or broad absorption lines (BALs), accurately quantifying
the noise, and determining the unabsorbed continuum baseline.
The primary purpose of this paper is to present a data set for which all of
the above steps have been implemented, drawing on the detailed
internal investigations by the BOSS collaboration, so that users can
easily perform their own \lya\ forest analyses.  
Our quasar continua predict the intrinsic quasar flux with errors
at the $\lesssim 5\%$ root-mean-squared (rms) level. For each spectrum, we introduce a
pixel-level mask to flag regions that may be affected by data
reduction problems, sky emission lines, DLAs, BALs, and non-\lya\ 
absorbers. 
The pipeline noise estimates are known to underestimate the true noise in the spectra
by up to $15 \%$ at wavelengths relevant to most of our \lya\ forest data ($3600\;\ang \lesssim \lambda \lesssim 5500\;\ang$); 
we include
corrections to remove these biases in the estimated pipeline noise.
This sample thus removes or corrects for the most obvious systematics
that might affect a \lya\ flux correlation analysis, although these
must be assessed in more detail in the context of any particular study.

The \citet{busca:2012} and \citet{slosar:2012} papers each employ
their own data selection criteria and quasar continuum treatments for their
primary BAO measurements.  However, an additional purpose of the
present study is to provide a fiducial sample and continuum fit
that can be used to compare the results from different methods.
Both papers therefore present additional BAO measurements for 
 the \lya\ forest sample and continua presented here.

Our sample is comprised of 
54,468 BOSS spectra 
that probe the \lya\ forest in the redshift range $2.0 < \zalp < 5.7$ (where $1+\zalp = \lambda/1215.67\;\ang$) at a typical 
sky area density of $\sim 16$ sightlines per square degree \citep{ross:2012}.
The co-moving volume encompassed by these sightlines is 
\beq
V =  c \iint \frac{ (1+z)^2 \;d^2_A(z)}{H(z)}\, {\rm d\Omega\; d}z \approx 20\; h^{-3}\, {\rm Gpc^3},
\eeq
where $\Omega$ is the solid angle, $H(z)$ is the Hubble expansion parameter, $d_A$ is the angular diameter distance, 
and we have taken the integral over the redshift range $2 < z < 3.5$ assuming a $\Lambda$CDM universe with 
$\Omega_{\Lambda}=0.7$, $\Omega_M = 0.3$, and $H_0 = 70\;\kms\,\mathrm{Mpc}^{-1}$, 
consistent with WMAP 7-year results \citep{komatsu:2011}.

This paper is organized as follows: \S~\ref{sec:boss_summary} summarizes the BOSS survey and 
provides relevant technical references; \S~\ref{sec:selection} presents the basic selection of suitable 
spectra from the overall BOSS quasar sample; \S~\ref{sec:products} describes the per-spectrum products
such as continua, masks, and corrections. We then describe several systematics of which users need to be aware
(\S\ref{sec:systematics}), before providing information on data access and usage (\S\ref{sec:data}).

%% file: boss_summary.tex
\section{Summary of BOSS Spectra}\label{sec:boss_summary}

BOSS \citep{dawson:2012} is one of four spectroscopic surveys\footnote{BOSS, SEGUE-2, MARVELS, and APOGEE; see \citet{eisenstein:2011}} in SDSS-III
\citep{eisenstein:2011} conducted on the 2.5-meter Sloan telescope \citep{gunn:2006}
at Apache Point Observatory, New Mexico.
The target selections in all these surveys were largely based on the SDSS imaging \citep{fukugita:1996,pier:2003,gunn:2006} 
that was completed in SDSS DR8 \citep{aihara:2011}.
The BOSS spectra are obtained by twin spectrographs inspired by
the original SDSS spectrograph design \citep{smee:2012}, that were 
completed in 2009 with improved volume phase holographic gratings, new CCDs,
more fibers, and smaller fiber diameter relative to the SDSS instruments.
The improvements produced roughly a factor of two increase in
instrument throughput and roughly a factor of two decrease in sky background, enabling studies
of a larger number of faint galaxies and quasars than what was possible in SDSS.
Both spectrographs separate the light into a blue and a red camera, covering the wavelength range of 
361 nm -- 1014 nm with a resolving power $\lambda/\Delta\lambda$ ranging
from 1300 at the blue end to 2600 at the red end.

As described in \citet{dawson:2012}, a typical plate is designed with 80 ``sky'' fibers
assigned to locations with no detected objects from SDSS imaging to provide an estimate of the sky background.
In addition, each plate includes 20 ``standard star'' fibers that are assigned to objects photometrically classified as F stars to 
calibrate the spectral response of the instrument.
About 160--200 fibers (40 deg$^{-2}$) are assigned to quasar candidates to probe
neutral hydrogen via absorption in the \lya\ forest.
The photometric classification and selection of quasar candidates for BOSS
spectroscopy produces 15--18 $z>2.15$ quasars deg$^{-2}$ \citep[see][]{ross:2012}.

Exposure times for each plate are determined during observations to obtain a uniform
depth across the survey;
on average, a plate is observed for five individual exposures of 15 minutes each.
The data are processed and calibrated by a data
reduction pipeline referred to as ``idlspec2d'' \citep{schlegel:2012}.
The functions of idlspec2d that are of consequence to \lya\ studies occur primarily in the
first stage of the pipeline, where data are extracted from the CCD images.
In this stage, the variance for each pixel is estimated using
read noise and the observed photon counts, sky background is subtracted using a model
derived from the sky fibers,
and flux calibration is performed using the spectra from the standard stars.
Each exposure produces a sample of independent, flux-calibrated spectra for
each object on the plate.
These spectra are wavelength sampled corresponding to the native CCD row spacing,
which can vary from exposure to exposure due to flexure and focus changes.
For each object, the individual flux calibrated spectra from each exposure
are compared to the ``primary'' spectrum with the highest signal-to-noise ratio.
A low-order polynomial is derived to provide a wavelength
dependent flux correction of each individual
spectrum to match the spectrophotometry of the primary exposure.
Finally, the individual spectra
are combined into a single spectrum that is binned into vacuum
wavelength pixels of $\Delta \log_{10} (\lambda)= 10^{-4}$, i.e. $\Delta v = 69.02 \kms$.
Each co-added spectrum is automatically redshifted and classified in the
final stage of idlspec2d \citep{bolton:2012}.

A spectrum of an object is identified by its plate, fiber number, and the
modified Julian day (MJD) of the last exposure contributing to the coadd. A small number of objects
have been multiply-observed\footnote{Most notably plate 3615 and 3647, which have together covered the same 7.1 deg$^2$ field
6 times in DR9}, and each have multiple spectra with different plate-MJD-fiber combinations.
SDSS-III Data Release 9 \citep[DR9;][]{sdss:2012} makes available these spectra
as one FITS-format file per plate-MJD-fiber (with the file prefix ``spec''), enabling
re-distribution of the exact subset of the spectra used for a particular
analysis or catalog.  The full version
of these files includes both the coadded spectrum and the individual
exposure spectra; the ``lite'' version does not include the individual exposures.
The format of these files is described in detail within the SDSS-III website\footnote{\url{http://data.sdss3.org/datamodel/files/BOSS\_SPECTRO\_REDUX/\allowbreak{}RUN2D/\allowbreak{}spectra/PLATE4/spec.html}}.
Header Data Unit (HDU) 1 of these files contains vectors with the vacuum wavelength solution (in logarithmic units),
co-added observed flux density (in units of $10^{-17} \fluxergs$), estimated inverse variance of the noise,
and bit mask vectors --- these are listed in the top half of Table~\ref{tab:spec_products}.
The spectra released (labeled with the file prefix ``speclya'') with this paper expand this format to include
additional masks, noise corrections, Damped \lya\ (DLA) system corrections,
and a continuum fit as described in \S~\ref{sec:products}.  Only HDU 1 is changed; 
other HDUs are the same as the original DR9 files.

\begin{table*}
\centering
\caption{Spectral Products in HDU 1 of `speclya' Product}
\begin{tabular}{ p{0.15\textwidth}  p{0.65\textwidth} }
\tableline
\multicolumn{2}{c}{Standard Pipeline Products} \\
\tableline
\verb|FLUX|		&	Coadded and calibrated flux density in units of $10^{-17} \fluxergs$ \\
\verb|LOGLAM|		&	Logarithm of wavelength in angstroms \\
\verb|IVAR|		& 	Inverse variance of flux \\
\verb|AND_MASK|	 	& 	AND mask\tablenotemark{a} \\
\verb|OR_MASK|		& 	OR mask\tablenotemark{a} \\
\verb|WDISP|	 		& 	Wavelength dispersion in dloglam units \\ 
\verb|SKY| 			& 	Subtracted sky flux density in units of $10^{-17} \fluxergs$ \\ 
\verb|MODEL|		&	Pipeline best model fit used for classification and redshift\tablenotemark{b} \\ \\
%\vspace{0.5em}
\tableline
\multicolumn{2}{c}{Value-Added Products} \\
\tableline
\verb|MASK_COMB|   	& 	Combined mask incorporating pipeline masks, sky-line masks, and DLA masks \tablenotemark{c} \\
\verb|NOISE_CORR|	& 	Pipeline noise corrections \\
\verb|DLA_CORR|	& 	Flux corrections for known DLAs \\ 						
\verb|CONT|		& 	Estimated quasar continuum in $1040-1600\;\ang$ restframe, in units of $10^{-17} \fluxergs$\\
\tableline
\end{tabular}
\tablenotetext{1}{See \url{http://www.sdss3.org/dr9/algorithms/bitmask\_sppixmask.php} for detailed description of the 
BOSS spectrum bitmask system}
\tablenotetext{2}{See \citet{bolton:2012}}
\tablenotetext{3}{See Table~\ref{tab:masks} for listing of combined masks}
\label{tab:spec_products} 
\end{table*}

%% file: selection.tex
\section{Sample Selection} \label{sec:selection}

\begin{table}
\centering
\caption{Selection Cuts for \lya\ Forest Sample}
\begin{tabular}{l  r   }
\tableline 
\tableline \\ [-1.3ex]
{\bf Description}    	& {\bf Number of Spectra}\\ [0.7ex]
\tableline \\ [-1.3ex]
DR9Q Quasars   & 87,822  \\  [0.2ex]
$\zq < 2.15$      & $-25,891$ \\  [0.2ex]
BAL quasars     & $-5,848$ \\  [0.2ex]
Low SNR          & $-924$   \\  [0.2ex]
Too many masked pixels & $-170$ \\  [0.2ex]
Negative continuum & $-521$ \\  [0.3ex]
\tableline \\ [-1.3ex]
Total   & 54,468 \\

%\vspace{0.5em}
\end{tabular}
\label{tab:cuts} 
\end{table}

In this section, we describe the spectrum-level cuts in order to select a useful sample of
\lya\ forest spectra from the overall BOSS DR9 sample. 

\begin{figure}
\epsscale{1.15}
\plotone{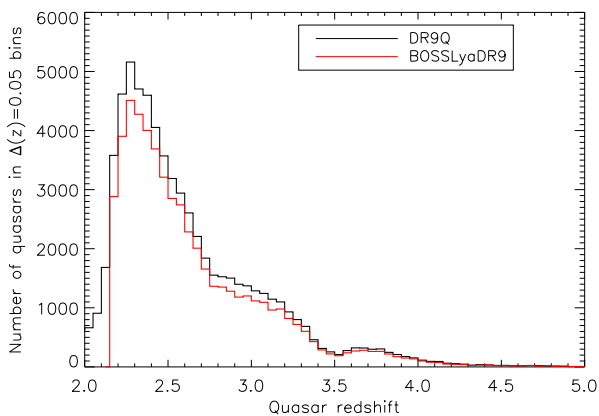}
\caption{\label{fig:zhist} Redshift distribution of high-redshift ($\zq > 2$) quasars in DR9Q, and in the present 
\lya\ Forest Value-Added Sample. The axes of this figure excludes
 22,617 DR9Q quasars with $\zq < 2$ and 22 quasars with $\zq > 5$. 
}
\end{figure}

We use as a parent catalog the BOSS DR9 quasar catalog of 87,822 objects
 visually confirmed as quasars \citep[][hereafter DR9Q]{paris:2012}. 
In addition to identifying quasars from the targeted candidates and flagging artifacts in the data, 
the visual inspection process of DR9Q also provides a visual refinement of the pipeline redshift estimates as well as 
identification of broad absorption line (BAL) quasars and damped \lya\ (DLA) absorbers.
The redshift distribution of the $\zq > 2$ quasars is shown in Figure~\ref{fig:zhist}, 
where we have adopted the visual inspection redshift,  \verb|Z_VI|, as the quasar redshift
(this definition is used throughout the paper unless noted otherwise). 
DR9Q lists only unique quasars; in the case of quasars that have multiple spectra, 
the catalog lists only the spectra 
(i.e. plate-MJD-fiber combination) with the highest signal-to-noise ratio (SNR).

\begin{figure}
\epsscale{1.15}
\plotone{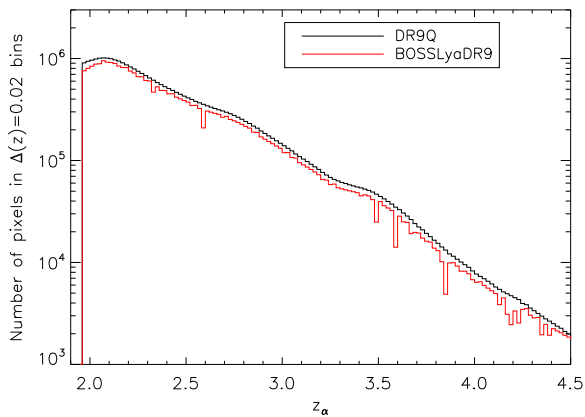}
\caption{\label{fig:pixelhist} Absorber redshift distribution of \lya\ forest pixels ($1041-1185 \ang$ restframe)
 in BOSS DR9. The black histogram shows all nominal \lya\ forest pixels from DR9Q, while the red histogram
 shows the final distribution in the present \lya\ forest sample, with the pixel-level masks applied
 (see \S~\ref{sec:masks}). The sharp dips in the distribution of the pixels represent
pixels which have been masked due to sky lines.
}
\end{figure}

It is clear from the Figure~\ref{fig:zhist} that the BOSS quasar target selection
\citep{ross:2012,bovy:2011} has selected an unprecedented number of high-redshift ($\zq>2$) quasars
with accessible \lya\ forest.
In principle, the minimum useable quasar redshift is that at which the 
quasar restframe \lya\ redshifts past the $3600\;\ang$
blue-end cutoff of the BOSS spectrograph, $\zq> 1.96$.
The absorber redshift distribution of all nominal \lya\ forest pixels in DR9Q is illustrated by the black histogram
in Figure~\ref{fig:pixelhist}.
However, for \lya\ forest analysis we want to ensure that each sightline contains
a reasonable number of \lya\ forest pixels in order to allow stable continuum fitting,
and cross-checks involving line-of-sight fluctuations.
We therefore set the minimum quasar redshift to $\zq \geq 2.15$:
this ensures at least $\npix  \approx 157$ useable \lya\ forest
pixels (corresponding to a minimum velocity pathlength of $\Delta v = 10800\;\kms$) 
in each sightline\footnote{Where $\npix = (\log_{10} \lambdamax - \log_{10} \lambdamin)/ 10^{-4}$,
 $\lambdamin=3600\;\ang$ is the nominal BOSS blue-end cutoff, and 
$\lambdamax= 1185 \times (1+2.15) = 3733\; \ang$ is set by the red-end of the quasar \lya\ forest
region.}. This criterion excludes less than 0.9\% of all possible \lya\ forest pixels, 
which are in any case from the noisy blue-end of the BOSS spectrographs, 
and hence carry less weight in any analysis.
The resulting pixel distribution is illustrated by
the red curve in Figure~\ref{fig:pixelhist}, although this also includes pixel-level cuts (\S~\ref{sec:masks}).
For consistency with the SDSS \lya\ forest analysis of \citet{mcdonald:2006}, we have defined the 
\lya\ forest region in each sightline to be $1041-1185\; \ang$ in the quasar restframe.
This range conservatively avoids the quasar proximity zone at the red-end and the 
quasar \lyb\ emission line at the blue-end. 

\begin{table*}
\centering
\caption{Description of BOSSLyaDR9 Catalog}
\begin{tabular}{ l c l }
\tableline 
\tableline
Column       & Format     & Description \\ 
\tableline 

\verb|SDSS-NAME|  & A19 & SDSS-DR9 designation  \\
\verb|RA |            &  F11.6 & Real ascension (J2000) \\
\verb|DEC     |     &  F11.6 &  Declination (J2000) \\
\verb|THING_ID |  & I10    &   Unique identifier  \\
\verb|PLATE |       & I5      &  Plate number  \\
\verb|MJD  |        &  I6      &  Spectroscopic MJD  \\
\verb|FIBER|         &  I5       &  Fiber number \\
\verb|Z_VI |      & F9.4     &  Visual inspection redshift from DR9Q \\
\verb|Z_PIPE |    & F9.4     &  BOSS pipeline redshift \\
\verb|SNR |         & F9.4      &  Median SNR ($1268-1380\;\ang$ rest) \\
\verb|SNR_LYA |   & F9.4       & Median SNR ($1041-1185\;\ang$ rest) \\
\verb|CHISQ_CONT| & F9.4      & Reduced chi-squared of continuum fit ($1216-1600\;\ang$ rest) \\
\verb|CONT_FLAG|    &  I2      & Continuum visual inspection flag \\
\verb|CONT_TEMPLATE|  & A8   & Quasar template used \\
\verb|Z_DLA |       &  F9.4     &  DLA absorption redshift \\
\verb|LOG_NHI |  &  F9.4       & Logarithm of DLA \ion{H}{1} column density in $\mathrm{cm^{-2}}$ \\
%\vspace{0.5em}
\tableline
\end{tabular}
\label{tab:cat} 
\end{table*}

In addition, broad absorption line (BAL) troughs may affect our continuum fitting and possibly introduce  
intrinsic quasar absorption into the \lya\ forest region. Therefore, we discard the 5,848 quasars visually
flagged as BAL quasars (\verb|BAL_FLAG_VI| = 1) in DR9Q. % 7532 or 5848???

Since our continuum-estimation technique uses the $1030 \ang < \lambrest < 1600 \ang$ range in the
quasar restframe spectrum, we also discard spectra in which more than 20\% of the pixels within this region are masked
by the pipeline (see \S\ref{sec:pipemask}). Similarly, we require that no more than 20\% of pixels within the $1041 \ang 
< \lambrest < 1185\; \ang$ \lya\ forest region are masked by the pipeline (see \S~\ref{sec:masks}). 

\begin{figure}
\epsscale{1.15}
\plotone{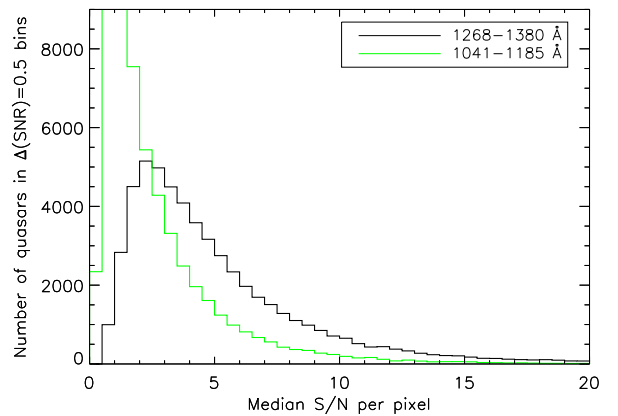}
\caption{\label{fig:snrhist} Signal-to-noise ratio distribution of the spectra in our \lya\ forest sample, 
evaluated both in the \lya\ forest region (green) and
redwards of quasar restframe \lya\ (black).
Note that there are 675 spectra with \snr $> 20$ per pixel over $1268-1380\;\ang$. 
}
\end{figure}

Next, we make a cut based on the SNR of the spectra. 
While the SNR requirements for 3D \lya\ forest flux correlation analysis are modest 
\citep{mcdonald:2007, mcquinn:2011}, it is difficult to estimate continua
from extremely noisy spectra. In the worst cases, even normalization is impossible. 
We therefore require our sample spectra to have a minimum median SNR of $\snr > 0.5$ per pixel 
evaluated over $1268-1380\; \ang$ restframe (redwards of the quasar \lya\ line) and a minimum median \lya\
forest SNR of $\snr > 0.2$ per pixel (after applying the noise corrections described in \S~\ref{sec:noisecorr}). 
We also cut spectra with more than one DLA (see \S~\ref{sec:dla}) within the \lya\ forest, but none
of the objects within the sample violated this criterion.
Spectra that have continua (see \S~\ref{sec:cont}) with negative regions are also discarded ---
this removes 521 objects that satisfy all other criteria in the sample, 
although these are all low-SNR spectra ($\snr < 1$ per
pixel in the forest).
In Figure~\ref{fig:snrhist}, we show the resulting median spectral SNR in our sample. 
Note that the effective SNR of the \lya\ forest region within each quasar spectrum 
is usually significantly lower than the red-side SNR due to IGM absorption and the increasing noise at the
blue-end of the BOSS spectrographs. 

Our final sample consists of 54,468 unique quasar spectra suitable for \lya\ forest analysis, with our cuts summarized 
in Table~\ref{tab:cuts} and the redshift distribution of all useable \lya\ forest pixels within our sample is shown by the
red histogram in Figure~\ref{fig:pixelhist} --- this also includes all pixel-level cuts described in subsequent sections of this paper. 
These objects are listed in a catalog, \verb|BOSSLyaDR9_cat| (available in both ASCII and FITS formats). 
The contents are summarized in Table~\ref{tab:cat}. 
The catalog and the individual spectra, described in the next section, can be 
downloaded from the SDSS-III website\footnote{\url{http://www.sdss3.org/dr9/algorithms/lyaf\_sample.php}}.

% Here we start discussing spectrum-level cuts

%% file: value_added.tex
\section{Per-spectrum Products}\label{sec:products}
In this section, we describe our expanded version of the BOSS 
high-redshift quasar spectra, intended to assist in \lya\ forest analyses.

We use as a starting point the per-object\footnote{These files are not strictly `per-object' as a small number of multiply-observed objects
have multiple plate-MJD-fiber combinations} `lite' co-add format released in SDSS DR9 \citep{sdss:2012}, which have the file prefix 
``spec''. 
The standard products packaged with this spectral
 format include the vacuum wavelength solution (in logarithmic units), 
co-added observed flux density (in units of $10^{-17} \fluxergs$), estimated inverse variance of the noise, 
and bit mask vectors --- these quantities are listed in the top half of Table~\ref{tab:spec_products}. 

However, a \lya\ forest analysis needs to take into account various systematics, e.g.
a detailed understanding of the pixel noise, masking of damped \lya\ absorbers (DLAs), and continuum fitting. 
In this section, we describe these additional products intended to assist in \lya\ forest analysis, 
which are composed of four primary components: (1) a continuum estimate for each quasar using the mean-flux regulated 
principal component analysis (MF-PCA) technique, (2) a noise correction to enable better noise estimates, 
(3) a simplified mask system to flag problematic pixels, and 
(4) corrections for intervening DLAs. 
These value-added products are packaged together with the original ``lite'' format products into new per-object spectra with 
the prefix ``speclya''.
While we have made it convenient to use the BOSS \lya\ forest data with this packaging, 
we emphasize we have not directly applied the new products unto the data, and users must perform the necessary operations themselves.

\subsection{Pixel Masks}
\label{sec:masks}

We now describe the bitmask system to flag pixels that should be discarded for 
\lya\ forest analysis. This process flags pixels identified by the pipeline as problematic, damped \lya\ absorbers (DLAs),
and sky emission lines. These mask bits are combined in a binary sense: e.g., a pixel in which
 bits 1 and 3 are set will store a value of $2^1 + 2^3 = 10$.
 These masks are stored in the \verb|MASK_COMB| vector in each spectrum, 
and the flags are summarized in Table~\ref{tab:masks}.

\subsubsection{Pipeline Mask}
\label{sec:pipemask}
The BOSS spectral pipeline \citep[\textsc{idlspec2d},][]{schlegel:2012} utilizes a system of 25 pixel mask bits to flag problems 
that may have occurred during the pipeline reduction process\footnote{\url{http://www.sdss3.org/dr9/algorithms/bitmask\_sppixmask.php}}. 
The \verb|ORMASK| vector in the co-added spectrum denotes pixels flagged by the pipeline in at least one of the 
individual exposures, while the \verb|ANDMASK| vector denotes pixels that were flagged in the equivalent
CCD column of all the individual exposures. The flagged pixels often have their inverse variances
set to zero by the pipeline, but the pipeline masks are more comprehensive.

In principle, all co-added pixels with \verb|ANDMASK| = 0 are free of problems, while flagged pixels may or may not
be useful depending on the user's application and discretion. However, in the DR9 version of the pipeline
mask bit 24 (``NODATA'', triggered by lack of detected flux) is erroneously set in the dichroic overlap region between 
the blue and red cameras, even when not all individual exposures were affected. 
This problem affects $9.7\%$ of all pixels, which are actually useable\footnote{Note that this issue affects only the
co-added spectra --- users of the individual exposures should \emph{not} ignore maskbit 24}. 

For simplicity, we amalgamate the pipeline \verb|ANDMASK| into our combined mask, 
such that maskbit 1 indicates pixels flagged by \verb|ANDMASK| (except \verb|ANDMASK| $=2^{24}$).

% Discuss number of pixels which are masked

\begin{table}
\centering
\caption{Combined Maskbits}
\begin{tabular}{l  c   p{0.27\textwidth} }
\tableline 
\tableline
{\bf Bit}    	& {\bf Binary}	& {\bf Description} \\
{\bf Name} & {\bf Digit}     &    \\
\tableline
\textsc{pipe} 	& 1 		&	Pipeline \textsc{andmask} is flagged \\
\textsc{sky}  	& 2		&	Improved mask for sky emission lines \\
\textsc{dla} 	& 3		& 	Mask for DLA cores \\
\tableline
%\vspace{0.5em}
\tablenotetext{0}{See \S~\ref{sec:masks} for full description}
\end{tabular}
\label{tab:masks} 
\end{table}

\subsubsection{Sky Mask} \label{sec:skymask}

 \begin{figure}
\epsscale{1.2}
\plotone{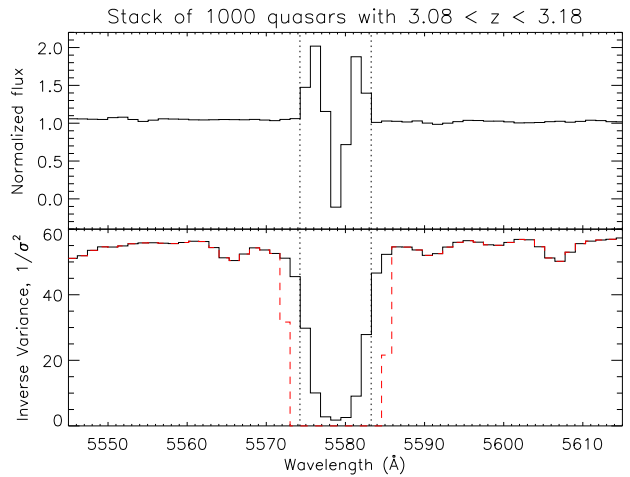}
\caption{\label{fig:sky_o1} Upper panel: Stacked flux from 1000 quasar spectra with $\zq = 3.08-3.18$
that have a flat intrinsic spectrum ($\lambrest \approx 1350 \ang$) around the $5577.338\ang$ \ion{O}{1}
sky emission line. The features are caused by increased noise variance from the sky line. 
The vertical dotted lines provide a visual reference point for the extent of the sky line's effect on the spectrum.
Lower panel: The pipeline noise inverse variance, where masked pixels have been set to zero using the pipeline
masks (black solid line) and our sky mask described in \S~\ref{sec:skymask} (dashed red line). The non-zero
pixels within the dotted vertical lines indicate that the pipeline masks do not adequately mask for the \ion{O}{1}
line, whereas our new sky mask has done so thoroughly. 
}
\end{figure}

\begin{figure*}
\plotone{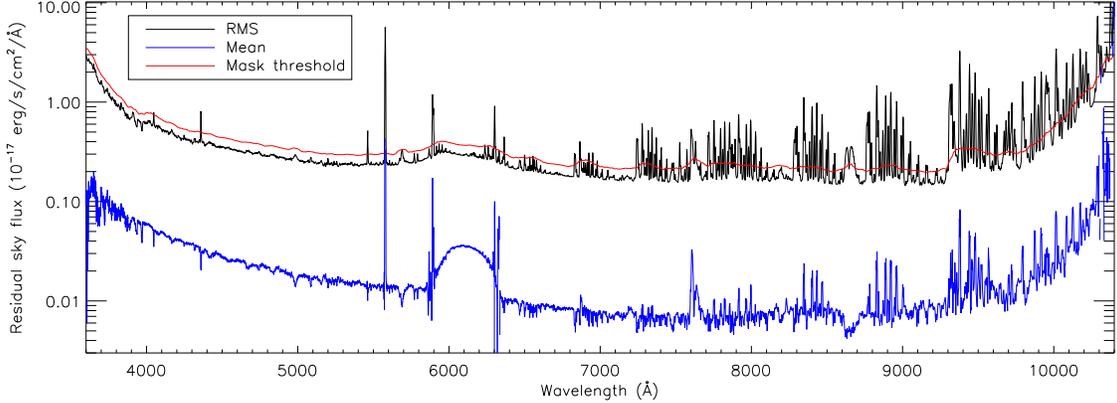}
\caption{\label{fig:skyresid} 
The mean (blue) and RMS (black) of the residual sky flux in the BOSS DR9 spectra, as estimated from sky fibers.
The red curve indicates the threshold used to define our new sky mask: wavelengths where the sky residual RMS rise
above this threshold are masked. The feature at $6000\;\ang \lesssim \lambda \lesssim 6300 \;\ang$ in the mean residual 
is an artifact in the dichroic region, where data from the blue and red CCDs overlap.
}
\end{figure*}

At the typical quasar magnitudes targeted by BOSS, the main contribution to pixel noise comes from the sky. 
This is particularly noticeable at pixels corresponding to the sky emission lines, where large deviations in flux are seen.
These pixels should be discarded since the astrophysical signal has been washed out by the sky variance.
In the pipeline, mask bit 23 (``SKYMASK'') is used to flag pixels where the object's estimated sky flux is 
(a) more than 10$\sigma$ above
the object flux, and (b) more than 1.25 times the median flux over the neighboring 99 pixels. 
However, we have found that using this criterion alone is insufficient
to fully mask strong sky emission lines --- this is illustrated in Figure~\ref{fig:sky_o1}, which shows the stacked spectrum
of 1000 quasars centered around the \waveion{O}{1}{5577.338} telluric emission line. 
The lower panel shows the corresponding inverse variances, with the pixels masked by the pipeline set to zero ---
the non-zero values within the envelope of the sky line indicate inadequate masking by the pipeline. 
In addition, weaker sky lines are often left unmasked by the pipeline.  

Since the sky calibration fibers in BOSS are themselves
processed by the standard pipeline --- including the sky subtraction estimated from all sky fibers in each plate ---,
the resulting residual spectra can be used to analyze the
efficacy of the latter procedure. 
The mean and rms of these sky residuals is shown in Figure~\ref{fig:skyresid}.
Using this, we generate a list of sky wavelengths to be masked as follows: we first define a `sky continuum' as the running
average of the residual rms fluctuation centered around a $\pm 25$ pixel window, and mask pixels
that are above $1.25 \times$ the sky continuum. The continuum and mask list are then iterated until they converge;
the final masking threshold is shown as the red curve in Figure~\ref{fig:skyresid}.
Pixels that are within 1.5 pixels of the listed wavelengths have mask bit 2 set in our combined mask, and should
be discarded in any analysis.
While there will still be a residual variance contribution from the sky subtraction, it should now vary smoothly with wavelength.  
The effect of the new pipeline mask can also be seen in the red dashed line in Figure~\ref{fig:sky_o1}: the \ion{O}{1} feature is now fully
masked. 

The mean residual of all sky-subtracted sky fibers is shown in the blue curve of Figure~\ref{fig:skyresid} --- there is a small
positive bias after the sky subtraction at the level of $\sim 0.01 \times 10^{-17}\, \fluxergs$, and rising to 
$\sim 0.1\times 10^{-17} \,\fluxergs$ at the blue and red ends of the spectra --- this is the cause of the zero-point
flux errors noted in Figure~4 of DR9Q.
This bias arises because the pipeline assigns a variance to the pixels based on their fluxes prior to the sky-subtraction step;
this underweights upwards sky fluctuations with respect to downwards sky fluctuations, providing an underestimate of the 
total sky background in low-SNR pixels. 
Since the flux transmission in the optically-thin \lya\ forest rarely drops to zero flux at BOSS resolution, we do not expect
this to be a significant issue in \lya\ forest analysis with BOSS data, but users studying DLAs and Lyman-limit systems (LLS)
need to take this into account.

\subsection{Noise Corrections} \label{sec:noisecorr}

\bfig
\epsscale{1.25}
\plotone{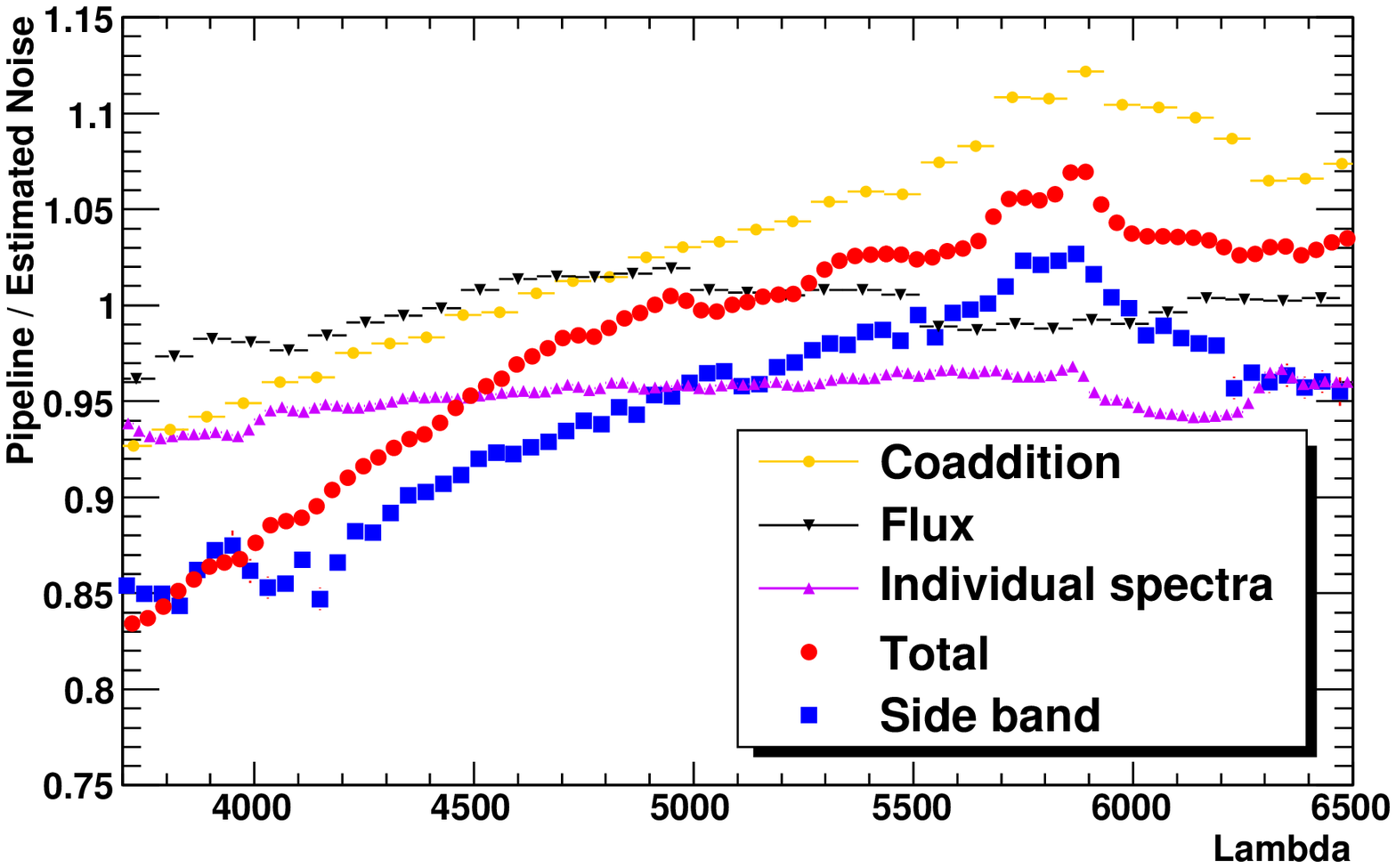}
\caption{\label{fig:noisetest}
The ratio of the pipeline noise estimate, $\sigma_{\rm p}$, to the actual flux dispersion in the spectra. 
The blue points denote this ratio as estimated from the quasar $1330\;\ang <\lambrest<1380\;\ang $ and 
$1450\;\ang <\lambrest <1500\;\ang$
side-bands. The purple points indicate the contribution to this bias from the pipeline estimates in the 
individual exposures that comprise each BOSS spectrum, while the yellow points show the bias introduced by the
pipeline co-addition procedure. The red points show the total correction from our procedure, ${\rm cor}_{\rm tot}$.
}
\efig

An estimate of the noise associated with each pixel in each spectrum, $\sigma_\mathrm{p}$, is provided by idlspec2d. 
However, the pipeline is known to suffer from systematic underestimates of the noise \citep[see, e.g.][]{mcdonald:2006, desjacques:2007} 
To investigate the extent of this, we examine the pixel variance in spectral regions that are intrinsically smooth and flat. 
We use two $\Delta \lambrest \sim 50\;\ang $ regions of quasar spectra (called `side-bands'), redwards of the \lya\ peak 
(so as not to be affected by absorption from the Ly$\alpha$ forest) and where the quasar continuum is relatively flat 
with restframe wavelength: $1330\;\ang <\lambrest<1380\;\ang $ and $1450\;\ang<\lambrest <1500\;\ang$.
For each individual quasar, we then compute the ratio of the pipe line error estimate, $\sigma_\mathrm{p}$, to the 
root-mean square (rms) of the flux dispersion about the mean within these side bands. 
This quantity is then averaged over all DR9Q quasars; with the varying quasar redshifts, this  
gives us a wavelength-dependent measure of the accuracy of the pipeline noise estimation 
(blue points in Figure~\ref{fig:noisetest}).
If the pipeline yields a perfect noise estimate, the plotted quantity should be unity at all wavelengths; on other hand, 
under (over) estimates will produce values below (above) unity.
The flux dispersion in the blue part of the spectra ($\lambda<4000\;\ang $) is seen to be about 15\% larger than expected from the 
noise estimate given by the pipeline. 
The discrepancy decreases with increasing wavelength, and the two estimates are in agreement at $\lambda \approx 5700\; \ang$. 

This test clearly indicates a wavelength-dependent miscalibration of the noise. 
However, since a fraction of the flux rms in the quasar side bands comes from intervening metals along the sightline, 
this procedure could be overly conservative in deriving the underestimation of the pipeline noise.  
Instead we recalibrate the pixel noise using three independent contributions derived from the data, 
which we shall now describe.  

Because the wavelength solution can vary between exposures, 
we first define a common wavelength grid with $2.5\;\ang$ pixels, about three times larger than on individual exposures. 
The flux $f$ in a given rebinned pixel is the weighted average of the flux of the contributing pixels of the original spectrum, 
with the weight taken to be the pixel inverse-variance $\sigma_\mathrm{p}^{-2}$. 
Input pixels which overlap two rebinned pixels are assigned to whichever rebinned pixel they overlap the most.
The correction terms ${\rm cor}_{\rm coadd} (\lambda)$, ${\rm cor}_{\rm exp}$ and $ {\rm cor}_{\rm flux} (\lambda, f)$ described below are computed from these rebinned single-exposure and coadded spectra. The total correction to the pixel noise is given by 
\begin{equation}
{\rm cor}_{\rm tot} (\lambda, f) =  {\rm cor}_{\rm exp} 
\times  {\rm cor}_{\rm coadd} (\lambda) \times {\rm cor}_{\rm flux}(f, \lambda) \;.
\label{eq:cortot}
\end{equation}

The various noise correction factors are:
\begin{description}

\item[Individual Exposure Correction, ${\rm cor}_{\rm exp}$]
We check the reliability of the pipeline error estimates on the individual exposures
that comprise each BOSS spectrum.
For instance, for $N$ exposures of an object, the distribution of the pull $S$ defined by
\begin{equation}
S = \frac{1}{\sqrt{N/2}} \times \sum_{i=0}^{N/2} \frac{f_{2i+1}-f_{2i}}{\sqrt{\sigma_{\mathrm{p},2i+1}^2 + \sigma_{\mathrm{p}2i}^2}}
\label{eq:flxindiv}
\end{equation}
should be a Gaussian with zero mean and  $\sigma_S=1$. 
In case of an odd total number of exposures, the last one is arbitrarily dropped in the computation of S.  We calculate $\sigma_S$ for each quasar 
as the rms over the wavelength range of the \lya\ forest and use that as a per-quasar correction,  ${\rm cor}_{\rm exp} = 1/\sigma_S (\lambda)$.  
In Figure~\ref{fig:noisetest} we plot ${\rm cor}_{\rm exp} $ as a function of the average observer-frame wavelength of the \lya\ forest, 
binned over multiple quasars per wavelength bin.  
The results indicate an underestimation of the pixel noise by about 6\%, with a wavelength dependence of less than 3\%.

\item[Co-addition Correction, ${\rm cor}_{\rm coadd} (\lambda)$]
We examine the propagation of the noise estimate in the coaddition process by comparing the noise given by the pipeline on the coadded frame (variance $\sigma_{\rm p, coadd}^2$) to the noise computed from the weighted mean of the $N$ exposures that contributed to the coadd, with variance $\sigma_{\rm p, mean}^2$ such that
$\sigma_{\rm p, mean}^{-2} = \sum_{i=0}^{N}\sigma_{\mathrm{p},i}^{-2}$, where the \sigpipei\ here are not corrected by ${\rm cor}_{\rm exp}$ 
since we assume that the noise estimate errors in individual exposures and those introduced by the co-addition process are orthogonal. 
The correction term for the co-addition process is defined by
$ {\rm cor}_{\rm coadd} = {\sigma_{\rm p, coadd} }/ {\sigma_{\rm p, mean}}$. This
increases with wavelength, from about $0.95$ at $\lambda=4000\ang$ to about 1.10 at $\lambda=6000\ang$, 
and is shown as the yellow points in Figure~\ref{fig:noisetest}.

\item[Flux-dependent Correction, ${\rm cor}_{\rm flux}(f, \lambda)$]
Within a given side-band, the ratio of the pixel noise, corrected by ${\rm cor}_{\rm coadd} \times {\rm cor}_{\rm exp}$, to the flux dispersion in the same rest-frame wavelength range for all quasars exhibits a flux dependence. We correct for this effect by applying a linear correction ${\rm cor}_{\rm flux}(f, \lambda)$, that we fit separately in five distinct wavelength bins, with the corrections bounded at ${\rm cor}_{\rm flux} \geq 0.9$. For typical fluxes in the Ly-$\alpha$ forest, the correction ranges between 1-5\% for $\lambda<5000\;\ang$ and up to 9\% for $\lambda>5500\;\ang$.
This mean over the spectra in our sample is shown as the black points in Figure~\ref{fig:noisetest}.

\end{description}

The pipeline noise estimate is divided by the overall noise correction, $\sigma_{\rm cor} = \sigma_{\rm p}/{\rm cor}_{\rm tot} (\lambda, f) $, to yield  
a more accurate noise estimate. The average correction for our spectra is shown as the red points in Figure~\ref{fig:noisetest}.
The corrections for each object in our sample are stored in the \verb|NOISE_CORR| vector of the corresponding spectrum. 
We have derived the above corrections only for the blue side of the spectra, $\lambda \le 6300\; \ang$, 
which reaches up to $z_{\alpha} =4.18$ (see Figure~\ref{fig:pixelhist}), 
which comprises the vast majority of \lya\ forest pixels. 
Pixels with $\lambda > 6300\; \ang$ have their noise corrections set to unity, ${\rm cor}_{\rm tot}=1.0$, 
such that the pipeline noise remains uncorrected on the red side of the spectra.

Several caveats should be kept in mind regarding these noise corrections. 
Some of the error in the pipeline noise estimates arises from scatter in the broad-band fluxing of the individual exposures and
act as a covariance between the individual pixels. 
As such, our noise corrections do not take into account off-diagonal terms of this overall covariance.
We also note that there is an uncertainty of several percent regarding these noise corrections, e.g. the 'side-band' and
'total correction' curves in Figure~\ref{fig:noisetest} disagree by several percent although the overall wavelength-dependence is in 
good agreement.
However, 3D correlation analyses should not be sensitive to errors
in the noise estimate although 1D analyses will require a more careful approach than what we have presented here. 

We expect the pipeline noise estimates to be significantly improved when the new spectral extraction algorithm of \citet{bolton:2010}
is implemented in subsequent BOSS data releases.
Alternatively, \citet{lee:2012c} will describe a probabilistic method for accurate noise estimation that allows separation of photon-counting and CCD noise
components.

\subsection{Damped \lya\ Absorbers}\label{sec:dla}

The cosmological utility of the optically thin \lya\ forest ($N_{HI} \lesssim 10^{-17} \;{\rm cm^{-2}}$) relies on the fact that the absorption field 
is a weakly non-linear tracer of the underlying dark matter fluctuations. 
Damped \lya\ absorbers \citep[DLAs, see][for a review]{wolfe:2005}, although also caused by neutral hydrogen absorption
in the IGM, are collapsed objects that do not have the same correspondence with the large-scale density field. 
Moreover, each individual DLA causes
large damped absorption profiles that affect large swathes ($\Delta v \gtrsim 5000\; \kms$) of affected sightlines. 
It is thus preferable to remove DLAs from any analysis of the large-scale \lya\ forest, although note that 
it is impossible to detect and remove all DLAs from the data, especially in the noisier spectra.

In their early analysis of BOSS data, \citet{slosar:2011} had simply discarded sightlines that contained 
DLAs identified by visual inspection. This is a sub-optimal approach, since while approximately 10\% of all \lya\ forest
sightlines contain DLAs, only $\sim 10\%$ of the \lya\ forest pixels in each affected sightline are directly impacted by the DLA.
It would therefore be more economical to mask the saturated absorption cores of the DLAs, 
and correct for the effect of their broad damping wings in affected spectra. 

To deal with DLAs, we use a combination of three different methods, described in \citet{carithers:2012} to detect DLAs in the BOSS
quasar sightlines: visual inspection, Fisher Discriminant Analysis, 
template cross-correlation. 

\bfig
\epsscale{1.15}
\plotone{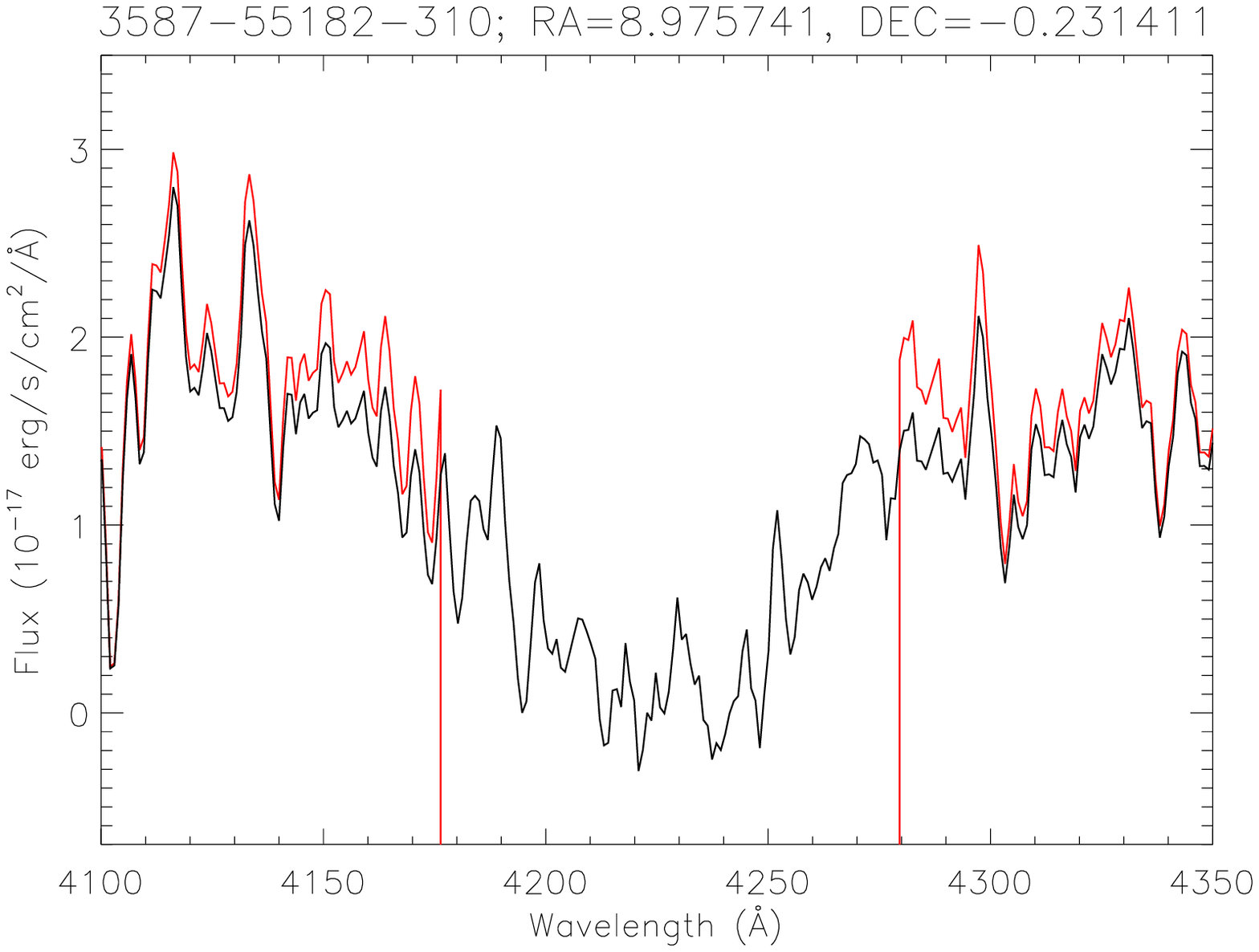}
\caption{\label{fig:dla}
The spectrum of a \lya\ forest sightline with a DLA at $z_\mathrm{dla}=2.477$, with a neutral hydrogen column density
 $\log_{10}{N_{HI}} = 21.19$. The red spectrum shows the same spectrum after applying the steps described in \S~\ref{sec:dla}:
 the central equivalent width $W$ (Equation~\ref{eq:ew}) of the DLA has been masked, while remaining pixels have been
 corrected for damping wings (Equation~\ref{eq:wings}). For clarity, both spectra have been smoothed with a 3-pixel mean boxcar function.
}
\efig

As mentioned above, all DR9Q spectra are visually inspected and spectra are flagged when a DLA is recognized by the inspector. 
In addition, we employ two automated procedures for identifying DLAs. The first, described in \citet{noterdaeme:2012},
 uses a set of DLA absorption profile templates of various column densities that are cross-correlated with the quasar spectra.  
If the correlation coefficient is sufficiently high, a fit to a Voigt profile is performed to measure the column density and DLA redshift. 
If associated metal absorption lines are present redwards of the quasar \lya\ emission line, they are used to refine the redshift. 
The second automated method, described in \citet{carithers:2012}, is based on a Fisher Discriminant \citep{fisher:1936} machine-learning algorithm. 
After an initial screening that identifies spectral regions that are consistent with zero flux density and inconsistent with the continuum, 
a fit to a Voigt profile is performed. The errors and chi-squares from the fit, along with the initial screening probability, are passed to a Fisher Discriminant that has been trained on the visual identification DLA sample. Metal lines, when present, are used by this method as well to refine the DLA redshift.

Any DLA recognition algorithm must balance the requirements for efficiency and purity, and the most severe challenge is in the regime of low SNR and low column density. Each of the three methods has strengths and weaknesses in this regard. To retain both high efficiency and high purity, we define a Òconcordance catalogÓ \citep{carithers:2012} consisting of all DLAs found by at least two of the three methods (in practice, the majority are found by all three techniques).
In those cases where a DLA is found by both
the template and FDA methods, the average of the two redshifts and column
densities is used. Both these methods have been tested on the same set of mock spectra \citep{font-ribera:2012a} that have 
DLAs artificially inserted; both yield detection efficiencies of $> 95\%$ for DLAs with $\log_{10} N_{HI} > 20.3$ in spectra with 
continuum-to-noise ratios\footnote{Where the continuum is, in this case, defined separately within each algorithm; see \citet{noterdaeme:2012} 
and \citet{carithers:2012} for details.} of $\mathrm{CNR} > 2$ per pixel.

For each DLA within this concordance catalog, we mask the wavelength region corresponding to the 
equivalent width  \citep{draine:2011}:
\begin{equation} \label{eq:ew}
W \approx \lambda_{\alpha} \left[ \frac{e^2}{m_e c^2} N_{HI} f_{\alpha} \lambda_{\alpha}
\left(\frac{\gamma_{\alpha} \lambda_{\alpha}}{c} \right) \right]^{1/2},
\end{equation} 
where $\lambda_{\alpha}=1216\; \ang$ is the rest-frame wavelength of the hydrogen 
\lya\ transition, $e$ is the electron charge, $m_e$ is the electron mass, $c$ is the 
speed of light, $N_{HI}$ is the \ion{H}{1} column density of the DLA, $f_{\alpha}$ is the
\lya\ oscillator strength, and $\gamma_{\alpha}$ is the sum of the Einstein $A$ 
coefficients for the transition. Pixels that are masked due to DLAs are flagged by maskbit 3 in our
combined mask.

Beyond this region, we correct for the damping wings of the DLA by multiplying each pixel in the spectrum with 
$\exp(\tau_\mathrm{wing}(\Delta \lambda))$, where
\begin{equation} \label{eq:wings}
\tau_\mathrm{wing}(\Delta \lambda) = \frac{e^2}{m_e c^3} \frac{\gamma_{\alpha} \lambda_{\alpha}}
{4 \pi} f_{\alpha} N_{HI} \lambda_{\alpha} \left(\frac{\lambda}{\Delta \lambda}\right)^2 
\end{equation}
and $\Delta \lambda \equiv \lambda - \lambda_{\alpha}$ is the wavelength separation 
in the DLA restframe.
Each of the spectra in our sample includes a vector, \verb|DLA_CORR|, that stores the damping wing corrections
$\epsilon_{\rm dla} \equiv \exp(\tau_{\rm wing})$;
this is set to unity in spectra without intervening DLAs. This correction vector should be multiplied into the
flux and noise vectors; alternatively, users might opt to make more stringent cuts based on the value of the damping
wing corrections.
Figure~\ref{fig:dla} shows a DLA in our sample, along with the masks and corrections that
we have applied to correct for it.

The \verb|Z_DLA| and \verb|LOG_NHI| fields in our catalog (Table~\ref{tab:cat}) lists the DLA absorber redshift and base-10 logarithm of the 
neutral hydrogen column density (in $\mathrm{cm^{-2}}$), respectively, for each spectrum in our sample. Both fields are set to $-1$ in spectra
where no DLAs are detected.

\subsection{Quasar Continua} \label{sec:cont}

In any \lya\ forest analysis, the transmitted \lya\ flux must be extracted by dividing
the observed flux by an estimate for the intrinsic quasar continuum. This is a non-trivial step even in high-SNR spectra. 
Traditionally, power-law extrapolation from $\lambrest > 1216\;\ang$ has
been used to estimate the quasar continuum in noisy spectra \citep[e.g.][]{press:1993}. However, this technique is now known 
to be unreliable due to a break in the quasar continuum at $\lambrest \approx 1200\ang$ \citep{telfer:2002}. 
Moreover, the uncertain blue-end spectrophotometry in BOSS (see \S~\ref{sec:specphoto}) 
makes continuum extrapolations highly unreliable. It is thus necessary to use the information
in the \lya\ forest itself to estimate the continuum.

For each BOSS DR9 quasar spectrum that satisfies our selection criteria in \S~\ref{sec:selection},  
we provide a continuum estimate using a modified version of the
mean-flux regulated principal component analysis (MF-PCA) technique described in \citet{lee:2012a}. 
This is technique essentially a two-step process: an initial PCA fit to the $\lambrest > 1216\;\ang$ region of the 
quasar spectrum to predict the \emph{shape} of the \lya\ forest continuum, followed by a 
`mean-flux regulation' step to ensure that the continuum \emph{amplitude} is consistent with
published constraints on the \lya\ forest mean-flux, $\fmean (z)$.

\bfig
\epsscale{1.15}
\plotone{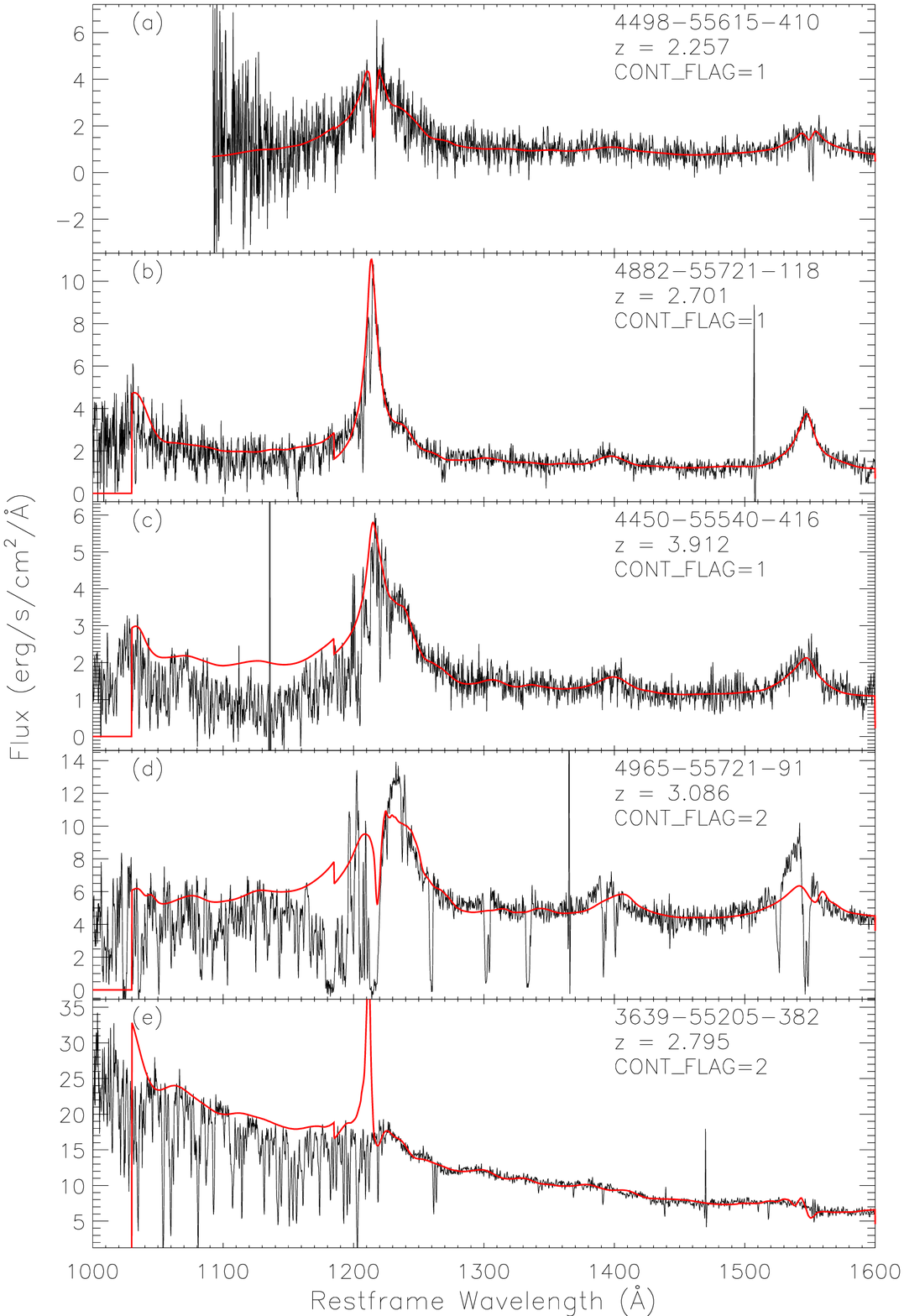}
\caption{\label{fig:cont_eg}
Spectra (black) of randomly-selected quasars from our sample, and their corresponding MF-PCA continua (red).
The two lower panels illustrate objects with inferior continuum fits ({\tt CONT\_FLAG}  = 2): in panel (d), 
a case where strong absorbers have stymied our efforts at absorption masking and biased the continuum fit; in panel
(d), a weak emission-line quasar that is not represented in our quasar templates.
These unsatisfactory continua comprise only 1.7\% of
the total sample.}
\efig

\subsubsection{PCA Fitting}

The first step in our continuum estimation process is to fit PCA templates to the quasar spectrum
redwards of its \lya\ emission line, in the $\lambrest = 1216-1600\;\ang$. 

However, since intervening metal absorption in that region might bias our continuum fit, we 
first execute a procedure to identify and mask these absorbers prior to fitting the continuum.
For this purpose, we follow the procedure described in \citet{lundgren:2009}.
First, we define a pseudo-continuum by using a variation of a moving mean that 
robustly fits both the quasar emission lines and flatter spectral regions over a broad range of quasar spectral morphologies. 
 Residual absorption features in the normalized spectrum are then each fit with a Gaussian to produce 
estimates of the equivalent width, $W$, and associated errors $\sigma_W$.  
Absorption lines detected with $W/\sigma_W \ge 3$ have their pixel inverse variances set to zero
and ignored in the subsequent steps\footnote{This absorber masking step was not done in 
\citet{lee:2012a} --- they instead used a iterative clipping method that was less effective in discarding
intervening absorbers}.

We obtain the initial PCA continuum, \cpca, by performing an inverse variance-weighted least-squares fit to the 
$1216\; \ang < \lambrest < 1600\; \ang$ region redwards of the quasar \lya\ emission line, 
using quasar templates with 8 principal components.
Two different PCA quasar templates were employed: 1) \citet{suzuki:2005} who used $z < 1$ quasars observed by the
Hubble Space Telescope, in which the $\lambrest < 1216\;\ang$ continuum can be clearly defined due to the lower
absorber density; and 2) \citet{paris:2011} who selected a sample of $z \sim 3$ quasars with high-SNR from 
SDSS DR7 and carried out spline-fitting on the \lya\ forest continuum to estimate the intrinsic quasar spectrum in that region.
Both templates are used to fit each BOSS quasar; the better-fitted template is then chosen based on the reduced chi-squared of the fit --- 
this is denoted by either `SUZUKI05' or `PARIS11' in the \verb|CONT_TEMPLATE| field of our catalog (Table~\ref{tab:cat}).
We find that for the DR9 sample, about 85\% of the quasars were better represented by the \citet{suzuki:2005} templates while
15\% were better-fit with the \citet{paris:2011} templates;
in contrast, the corresponding percentages in DR7 \citep[c.f.][]{lee:2012a} were 30\% and 70\%, respectively.
 We suspect that this is because fainter quasars are targeted in DR9 than in DR7; these faint quasars
better matched by the lower-luminosity quasars that comprise the \citet{suzuki:2005} templates.

However, not all the BOSS quasars are well-described by either of the quasar templates described above, 
in which case we cannot obtain a well-fitted PCA continuum. There are also cases in which strong absorption 
systems lying on top the quasar emission lines (most notably \lya) were not identified by the absorption-masking procedure, 
which biases the continuum fit.  
Initially, we attempted to use the reduced chi-squared
statistic, $\chi^2/\nu$, to quantify the fit quality, where $\nu=\npix-11-1$ is the number of degrees of freedom in our 11-parameter
PCA model and $\npix$ is 
the number of pixels evaluated in the range $1216 \;\ang < \lambrest < 1600 \;\ang$.
We found that while most objects with $\chi^2/\nu > 2$ 
were indeed badly-fitted, many unsatisfactory fits had $\chi^2/\nu \sim 1$, 
mostly in situations where absorption features were fitted by the principal components, giving unphysical continua.
We have therefore visually inspected all the fitted continua in the restframe region redwards of 
$1216\; \ang$, and flagged objects that were not well-fit by our PCA templates. 
We have listed both the reduced chi-squared and visual continuum flags in the 
\verb|CHISQ_CONT| and \verb|CONT_FLAG| fields, respectively, of the \verb|BOSSLyaDR9_cat| catalog.

Our convention for the visual inspection continuum flags is as follows:
\begin{description}
\item[{\tt CONT\_FLAG}=1] The fitted PCA continuum appears to describe the intrinsic quasar continuum well. 
We allow unphysical features in the continua (e.g.\ the `absorption feature' near $\lambrest =1216\;\ang$ in panel (a) of Figure~\ref{fig:cont_eg}), 
if they do not impact the overall fit.
Comprises $98.3\%$ of all spectra in our sample.
\item[{\tt CONT\_FLAG}=2] The fitted PCA continuum is badly fit and does not resemble the intrinsic quasar spectrum.
These cases tend to be caused by either very strong absorbers that have eluded our masking process, or quasars with continuum shapes 
that are not captured by our templates (see panels (d) and (e) in Figure~\ref{fig:cont_eg}).
These comprise $1.7\%$ of all spectra in our sample.
\end{description}
Because we apply the mean-flux regulation step (next section), even the worst continua with \verb|CONT_FLAG|=2
should yield rms continuum errors well under $\sim 10\%$. We therefore do not recommend that users discard 
spectra based on these flags, but use them as a possible systematic check.

\subsubsection{Mean-flux Regulation}

The initial PCA continuum fit, $\cpca$, provides a prediction for the shape of the weak quasar emission lines in the 
$1041\ang < \lambrest < 1185 \ang$ region, but the overall amplitude is uncertain 
due to the quasar power-law break and spectrophotometric errors.
We therefore require that each quasar continuum match the expected \lya\ forest mean flux evolution, 
given by \citet{faucher-giguere:2008a} --- we use their power-law-only fit without metal corrections:
\beq \label{eq:meanflux}
\langle F \rangle (z) = \exp[-0.001845(1+\zabs)^{3.924}],
\eeq
where $\zabs$ is the absorber redshift.

We fit a linear correction function of the form $(a + b \lambrest)$, 
such that the final continuum, \cmf,  yields a mean-flux in agreement with
Equation~\ref{eq:meanflux}. This is different from \citet{lee:2012a}, who performed this
step using a quadratic fitting function of the form $(1 + a \hat{\lambda} + b \hat{\lambda}^2)$, where
$\hat{\lambda} =  \lambrest/1280\ang - 1$ --- we changed to the linear correction function since it is easier
to compute analytic corrections for large-scale power along the line-of-sight \citep[e.g., Appendix A in][]{slosar:2011}. 

In addition, the weighting is carried out differently. In \citet{lee:2012a}, the correction function
was fitted to the \lya\ forest split into 3 restframe bins, with the weights in each bin given by the
inverse variance estimated through a 
bootstrap procedure; for our continua, we instead fit the correction function directly to the individual pixels, 
with weights given by the inverse of $\sigma^2 = \sigma_N^2 + \sigma_F^2$, where $\sigma_N^2$ is the
corrected (see \S~\ref{sec:noisecorr}) pipeline noise variance and
\beq
\sigma_F^2  (z)= 0.065 [(1+\zabs)/3.25]^{3.8} \langle F \rangle^2(z)
\eeq
is the intrinsic variance of the \lya\ forest within a $69\;\kms$ pixel, as estimated from the redshift
evolution of the power spectrum \citep{mcdonald:2006}; and $\langle F \rangle(z)$ is given by
Equation~\ref{eq:meanflux}.
In this fit, we use only pixels with $\lambda \ge 3625\;\ang$ in order avoid the regions most severely affected by 
the sky noise (c.f. Figure~\ref{fig:skyresid}).

The mean-flux regulation corrections are applied to the initial continuum estimate, $\cpca$, bluewards of $1185 \ang$. 
This introduces a discontinuity at $1185 \ang$ in the final continuum that is unphysical,
but we do not expect any practical issues to arise from this discontinuity if our assumed \lya\ forest range is adopted.
We have found that a small number ($\sim 500$) of extremely low SNR ($\snr \lesssim 1$) spectra
have continuum that go negative at some wavelengths. Since this situation is clearly unphysical, 
we therefore discard these quasars from the overall sample. 

For all 54,468 quasars in our sample, we provide estimated continua (in the \verb|CONT| vector of each file) that cover
 the quasar restframe range $1040-1600\ang$; the continua outside of this range are set to zero. 
From the tests on mock spectra by \citet{lee:2012},
we expect the typical rms error of the MF-PCA continua to be around 6\% at $\snr \sim 2$ per pixel (evaluated within the forest), 
dropping to $\sim 4\%$ at higher SNR ($\snr \gtrsim 5$ per pixel).

Several caveats must be kept in mind with regards to our continua. First, 
because the MF-PCA method requires an external constraint of the \lya\ forest mean flux
evolution, the continua presented here cannot be used to make an independent measurement
of the \lya\ forest mean flux --- they are primarily intended to provide a good per-pixel 
continuum estimate at the expense of zeroth order information on the mean flux. 
Another possible issue is that the mean-flux regulation removes
large-scale flux power along the line-of-sight, which means that our continua will not yield 
accurate measurements of one-dimensional flux power unless corrections are applied (A.\ Font-Ribera, private communication). 
In addition, the method would introduce some correlations in the continua in neighboring lines-of-sight.
Nevertheless, we do not expect this effect to bias measurements of the BAO peak position.

%% file: systematics.tex
\section{Known Systematics} \label{sec:systematics}
In this section, we describe several issues in the BOSS spectra that could have
an impact on cosmological analyses.

\subsection{Spectrophotometric Errors} \label{sec:specphoto}

\begin{figure}
\epsscale{1.15}
\plotone{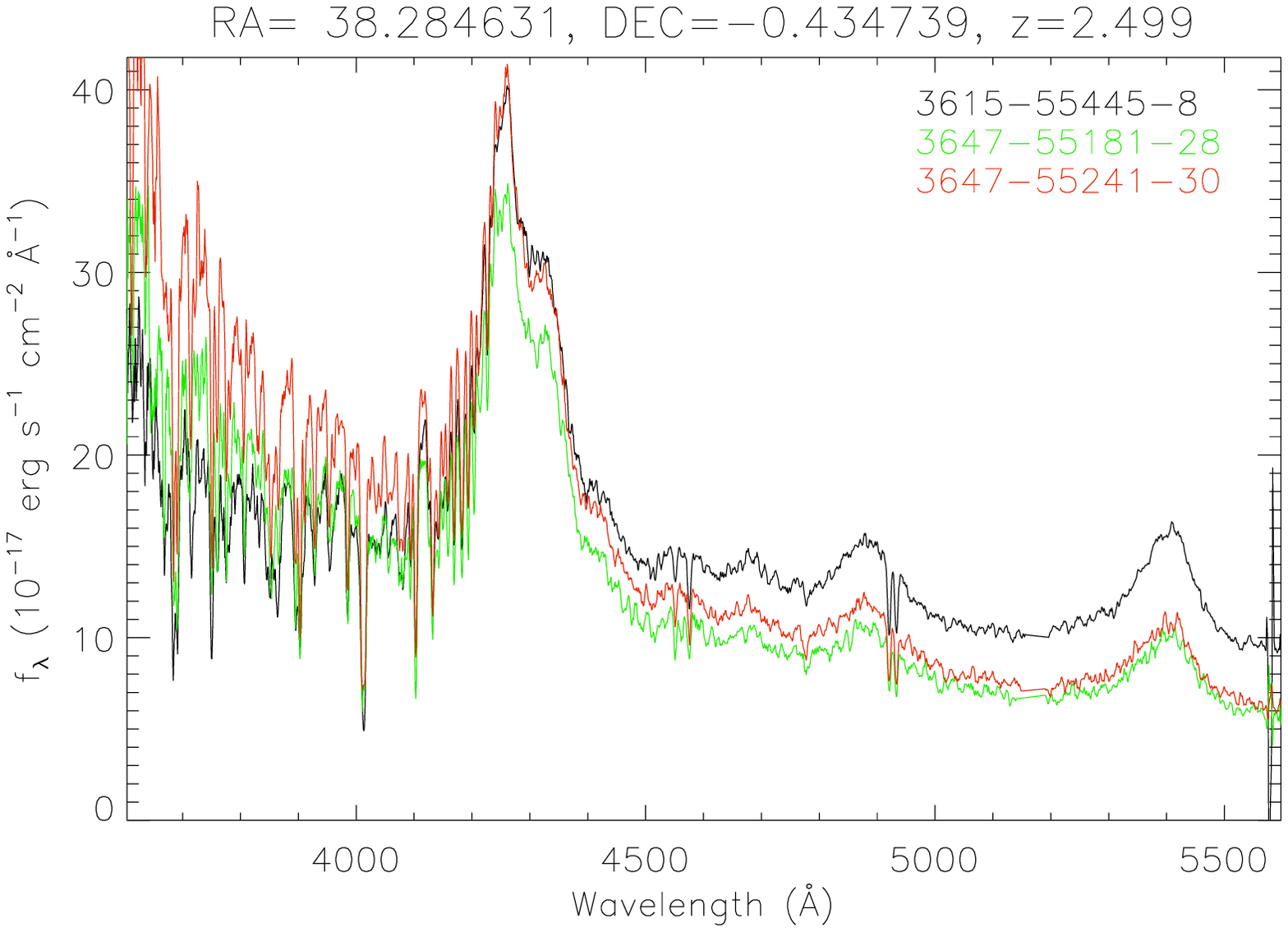}
\caption{\label{fig:multiobs_spec}
Multiple observations of the same BOSS quasar, illustrating the effect of differential atmospheric 
refraction on the spectrophotometry. The spectra have been smoothed with a 5-pixel boxcar function
 for clarity. The spectrum with the plate-mjd-fiber combination 3615-55455-8 is the `primary' 
 spectrum catalogued in DR9Q. 
This is an unusually bright BOSS quasar, with a magnitude $g = 18.66$ and $\snr \approx 23$ 
 per pixel.
}
\end{figure}

\begin{figure*}[t]
\plotone{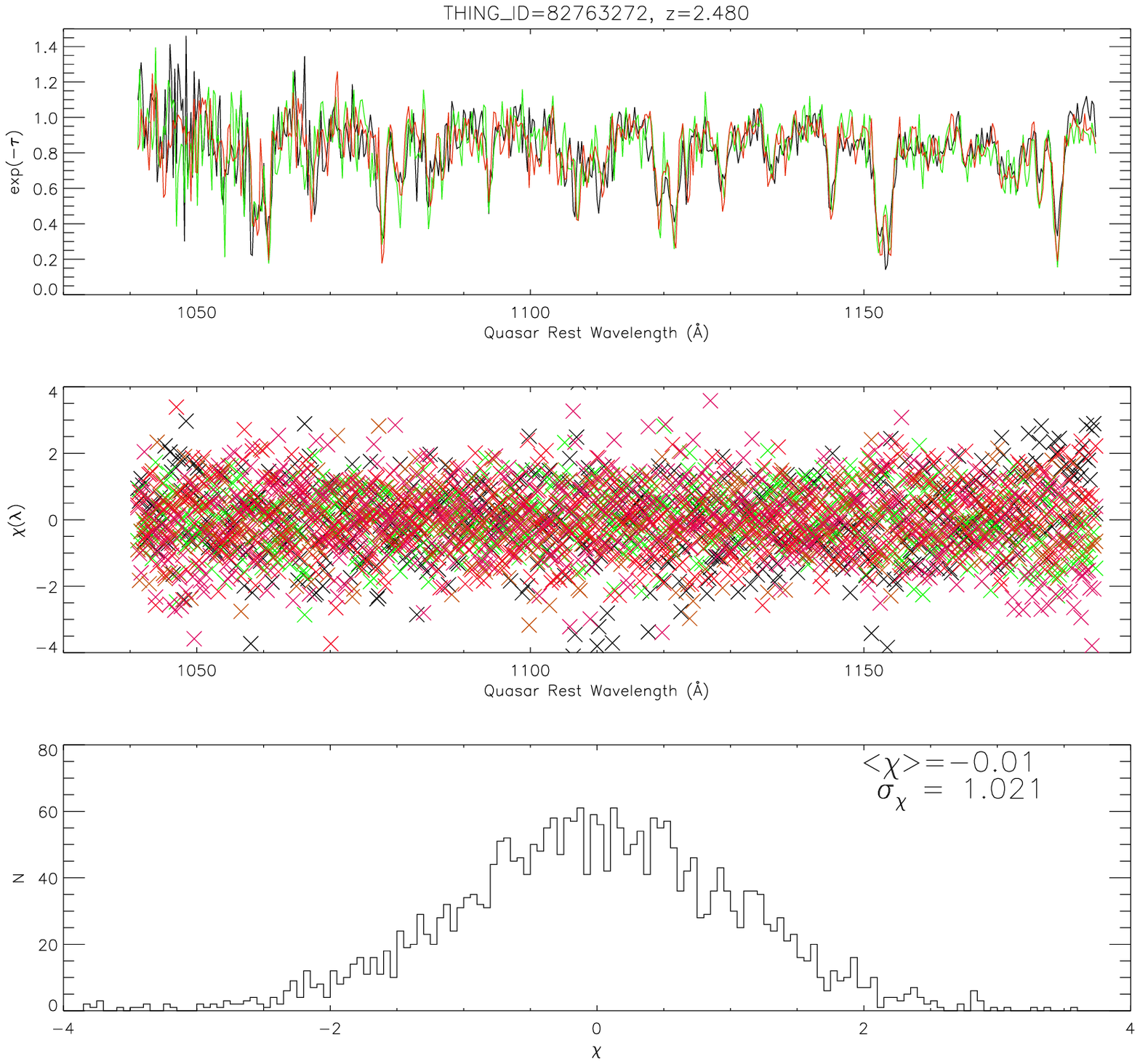}
\caption{\label{fig:multiobs_forest}
Top panel: A comparison of the extracted \lya\ forest transmission fields of the multiple-observations shown
in Figure~\ref{fig:multiobs_spec}, derived by applying the continuum estimation described in \S~\ref{sec:cont} on each individual 
observation. The transmission fields appear to be visually consistent with each other. 
Middle panel:  The pull distribution, $\chi(\lambda)$ (Equation~\ref{eq:cont_pull}), from the multiple
transmission fields in the top panel. 
Bottom panel: The histogram of the pull distribution shown in the middle panel. It is consistent with a 
Gaussian distribution with unit standard deviation, indicating our continuum-fitting method has removed the
spectrophotometric variations from the transmission field.
}
\end{figure*}

To improve the blue-end signal-to-noise for \lya\ forest analysis at $z \approx 2$, we have 
made the following modifications in the way that quasar fibers are attached to the plug-plates on the 
BOSS spectrograph: 
(a) thin ($175-300\; {\rm \mu m}$) washers were attached to the plate plug-holes to provide an axial offset, 
and (b) the positions of the quasar fibers are offset by up to $\sim 0.5''$ in order to maximize the light
entering the fiber when taking into account the atmospheric differential refraction (ADR) at the designed
plate hour-angle \citep{dawson:2012}. 
These adjustments shift the effective focus from $5300 \ang$ (as originally designed) to $\sim 4000 \ang$, 
which improves the blue-end signal-to-noise for \lya\ forest analysis. 
However, at time of writing the flux standard stars are observed only through fibers
without these offsets, 
rendering the spectrophotometric calibration highly uncertain on the blue end. 
A BOSS ancillary program is now in place to observe a number of spectrophotometric standard stars
through the quasar fibers in order to improve the spectrophotometric calibration, 
but the results of this program will not be incorporated until future data releases.

Furthermore, the blue end of the spectrum is more susceptible to differential atmospheric refraction, 
causing the spectrophotometry of the spectra to vary as a function of observed zenith angle. 
This effect is illustrated in Figure~\ref{fig:multiobs_spec}, where we show three spectra of a BOSS quasar that
had been observed on multiple nights. 
An important consequence of this uncertain spectrophotometry is that 
\emph{quasar continua cannot be directly extrapolated from redwards 
($\lambrest > 1216\;\ang$) of the quasar \lya\
emission line}, e.g., using a power-law.   
Direct extrapolation generally produces a large continuum error even in 
spectra with good flux calibration, but the existing spectrophotometric
errors in BOSS means that direct extrapolation will be biased on average
(see Figure~5 in DR9Q).

However, the MF-PCA continua included with this sample ameliorates 
the spectrophotometric errors. This effect is illustrated in Figure~\ref{fig:multiobs_forest}, 
where we compare the transmitted flux fields extracted from the multiple observations 
of same object shown in Figure~\ref{fig:multiobs_spec}, with MF-PCA continua fitted to each 
individual spectrum. One sees from the top panel that
the resultant flux fields appear consistent with each other, within the noise, despite the large differences in 
spectrophotometry as seen in Figure~\ref{fig:multiobs_spec}. 
We further quantify this by computing another form of the pull: 
\beq \label{eq:cont_pull}
\chi(\lambda) = \frac{F_i(\lambda)  - \hat{F}(\lambda)}{\sigma_{\mathrm{cor},i}(\lambda)/\cpca(\lambda)},
\eeq
where $F_i$ is the transmitted (i.e.\ continuum-normalized) flux from the different observations denoted by subscript $i$, 
$\hat{F}$ corresponds to the average of all the observations at a given wavelength, and $\sigma_\mathrm{cor}$ is the
corrected pipeline noise. The values of $\chi$ from the multiple observations are shown in the middle
panel of Figure~\ref{fig:multiobs_forest}. The bottom panel shows the combined histogram of all the
$\chi$ distributions, which appears Gaussian with a standard deviation 
close to unity, implying that 
pixel noise is sufficient to account for
the variance in the derived transmission fields and the variance from the spectrophotometric errors have been corrected. 
Although this particular quasar 
has unusually high \snr, we have shown that errors in the relative spectrophotometry do not significantly
affect our continuum estimates.  

\subsection{Flux Calibration Artifacts} \label{sec:fluxcalib}
\bfig
\epsscale{1.15}
\plotone{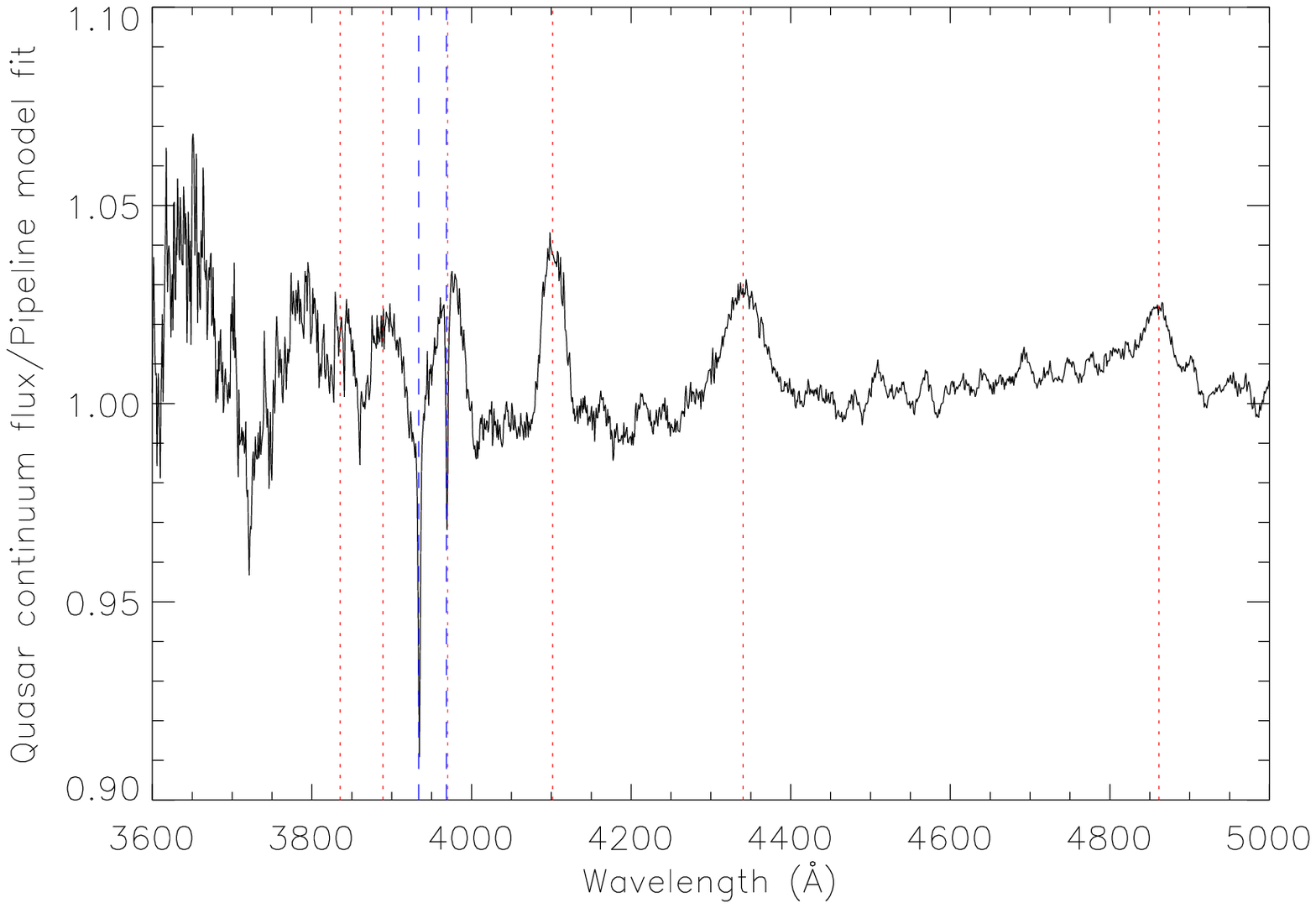}
\caption{\label{fig:fluxfeatures}
The ratio from dividing 28,848 quasars with $g < 20.5$ by their pipeline PCA 
models citep{bolton:2012}. The features correspond to Balmer line wavelengths (H$\delta$ through H-7 
are shown as red vertical dotted lines), while the prominent absorption lines at $3933.7\;\ang$ and $3968.5 \;\ang$ 
(blue vertical dashed lines) are possibly a consequence of \ion{Ca}{2} H\&K absorption by the interstellar medium.
}
\efig

We showed in \S~\ref{sec:skymask} that imperfect subtraction of prominent sky emission lines can lead to spectral artifacts if not carefully dealt with. However, imperfect flux calibration can also lead to artifacts. This conversion from counts to flux is achieved, in part, by placing fibers on F sub-dwarf stars and using them as spectrophotometric standards. The derived calibration vectors are largely fixed for all fibers plugged into each plate, fed to each of the two BOSS spectrographs. These vectors can be characterised as constant for fibers 1-500 and 501-1000 and therefore their flux calibration may vary from `half-plate' to `half-plate'. These spectrophotometric standards show pronounced Balmer absorption lines and these must be masked and interpolated over for accurate fluxing. There are potential systematic errors associated with this procedure as discussed in the DR2 and DR6 release papers 
\citep{abazajian:2004,adelman-mccarthy:2006}; these were ameliorated in the pipeline reduction of those releases but seem to have reappeared in the
DR9 spectra.

To illustrate these artifacts, in Figure~\ref{fig:fluxfeatures} we stack the ratio of the flux and the best-fit pipeline PCA model \citep{bolton:2012}
from all 28,848 good quasar
spectra in the DR9 sample where the observed spectroscopic r-band magnitude
was brighter than 20.5 (\verb|CLASS|=`QSO', \verb|ZWARNING|=0, \verb|SPECTROSYNFLUX[2]| $>$ 6.3 nMyg).
These ratios, and the formal pipeline errors, are combined at each
observer-frame (barycenter) wavelength using a weighted mean with 3-sigma outlier rejection.
We exclude any data points within $100\;\ang$ of 31 possible emission line locations at the quasar redshift,
blueward of \lya, or where the template flux density is lower than $ 0.5\;\mathrm{erg/s/cm^2}/\ang$.
These exclusions imply that only the smooth quasar continuum
at $\lambrest > 1216\;\ang$ contributes to the stack, while at $\lambda < 4000\;\ang$ only low-redshift
quasars at $z < 2.0$ contribute..

In the resulting ratio shown in Figure~\ref{fig:fluxfeatures}, we see unwanted wavelength dependent structure at the $\sim 2-3\%$ level. 
The prominent \ion{Ca}{2} H\&K absorption lines, at $3968.5 \;\ang$ and $3933.7\;\ang$ respectively are thought to be some combination of absorption by the solar neighbourhood, the interstellar medium and the Milky Way halo. In addition, artifacts are present at Balmer transition wavelengths due to imperfect correction of standard star absorption lines.

At time of writing, this issue has not yet been fully corrected in the BOSS pipeline, so users must take this effect into account in their analyses. 
As an interim solution, the ratio shown in Figure~\ref{fig:fluxfeatures} can be used as a correction vector and has been made publicly 
available with our sample (see \S~\ref{sec:data} for download instructions)  ---
the DR9 pipeline fluxes should be divided by this correction vector to remove the Balmer features, and other fluxing artifacts, 
on average.  This correction was applied to the spectra prior to the continuum fitting process in \S~\ref{sec:cont}, but it is not otherwise
incorporated into the fluxes in individual ``speclya'' spectra --- users need to carry out this procedure themselves.

It should be noted that \citet{busca:2012} find that the magnitude of these artifacts are comparable for the two BOSS spectrographs and that the square-root of the half-plate-to-half-plate variance is no larger then 20-100\% of the mean deviation (depending on the test applied). They conclude that the error introduced by half-plate-wide deviations from this correction vector is insignificant for their analysis.

%% file: conclusions.tex
\section{Data Access and Usage Guidelines} \label{sec:data}
The files associated with the BOSS DR9 \lya\ Forest Sample described in this paper can be downloaded from 
the SDSS-III website\footnote{\url{http://www.sdss3.org/dr9/algorithms/lyaf\_sample.php}}. 
We have generated \verb|BOSSLyaDR9_cat|, a catalog listing the objects in this 
sample along with the additional information useful for \lya\ forest analysis (described in 
Table~\ref{tab:cat}). It available in both FITS and ASCII formats.

The main components of the sample are individual `speclya' spectral files corresponding to each object in our sample.
These files are a value-added version of the `lite' per-object BOSS format 
(see \S~\ref{sec:boss_summary}),
but with additional masks and corrections as listed in Table~\ref{tab:spec_products}.
Note that these masks and corrections have \emph{not} been applied to the pipeline flux, \fpipe, nor inverse-variances, 
$\wpipe \equiv \sigpipe^{-2}$ by default, but are included as separate vectors in each file.
The flux correction described in \S~\ref{sec:fluxcalib} is available in a separate file, \verb|residcorr_v5_4_45.dat|, that can also be downloaded
from the aforementioned website.

For a standard analysis, users should use all objects listed by their unique plate-MJD-fiber
combination in the catalog, and each object will 
have a corresponding ``speclya'' spectrum file labeled by plate-MJD-fiber, grouped in subdirectories by plate number.
The \lya\ forest pixels in the range $1041\;\ang < \lambrest < 1185\;\ang$ should be selected from 
each spectrum in the catalog, where the quasar restframe is defined with respect to the redshift given by the \verb|Z_VI| 
(visual inspection redshift) field in the catalog. 
Pixels with zero inverse-variance or non-zero bits in the \verb|MASK_COMB| vector 
should then be discarded or masked. The pipeline flux, \fpipe\ (\verb|FLUX| in the speclya files), is then divided by the flux calibration 
corrections\footnote{Interpolated to the individual wavelength grids from {\tt residcorr\_v5\_4\_45.dat} described in \S~\ref{sec:fluxcalib}}, 
$\epsilon_{\rm flux}$ and 
multiplied by the DLA damping wing corrections, $\epsilon_{\rm dla}$ (\verb|DLA_CORR|), before dividing by the MF-PCA continua, 
$C_{\rm MF}$ (\verb|CONT|), to obtain the transmitted \lya\ forest flux.
The same operations are applied to the pipeline noise, \sigpipe\ (although this is stored as the inverse-variance, 
$\wpipe \equiv \sigpipe^{-2}$, \verb|IVAR| in the data files), but with the additional step of dividing by the 
noise corrections ${\rm cor}_{\rm tot}$ (\verb|NOISE_CORR|).

In other words, the \lya\ forest transmission field, $F_i$, is extracted from each spectrum $i$ like so: 
\beq
F_i(\zalp) = \fpipei(\lambda)\; \left( \frac{\epsilon_{{\rm dla},i}(\lambda)}{\epsilon_{\rm flux}(\lambda)\; C_{{\rm MF},i}(\lambda)}\right),
\eeq
where $(1+\zalp) = \lambda / 1215.67\;\ang$.

The corresponding inverse variance weights are derived from the pipeline inverse variances, \wpipei, 
as follows:
\beq
\wfi(\zalp) =  \wpipei(\lambda) \, {\rm cor}_{\rm tot}^2(\lambda) 
\left( \frac{\epsilon_{\rm flux}(\lambda)\; C_{{\rm MF},i}(\lambda)} {\epsilon_{{\rm dla},i}(\lambda)}\right)^2
 \eeq
All pixels with \verb|MASK_COMB| set or $\wpipei=0$ should be masked or discarded. 

\section{Conclusions} \label{sec:conclusions}

We present the public release of the BOSS DR9 \lya\ Forest Sample, 
a set of 54,468 spectra suitable for \lya\ forest analysis selected from the BOSS DR9 quasar catalog, 
taking into account criteria such as redshift, SNR, and quality of spectra.
For each spectrum, we also provide the following products designed to aid in \lya\ forest analysis:
\begin{itemize}
\item A simple maskbit system to flag pixels that may be affected by pipeline artifacts
or sky emission lines, or that lie within DLA cores.
\item Corrections for DLA damping wings.
\item Noise correction vectors to make the pipeline noise estimate consistent with the actual pixel dispersions.
\item An MF-PCA continuum estimate accurate to $5\%$ rms at the median \snr\ of the data.
\end{itemize}

In addition, we have also discussed two sytematics in the data that may affect \lya\ forest analyis. 
The relative spectrophotometry is uncertain due to steps in the observational procedure
taken to boost the \lya\ forest SNR, 
but we argue that the MF-PCA continua provided here removes these effects to first-order.
We also discuss artifacts in the spectra caused by the errors in the flux calibration, and provide a global correction as
an interim solution prior to a more thorough solution within the BOSS pipeline.

While this sample is a convenient resource for users intending to work with the BOSS \lya\ forest
data, we encourage users to make their own decision on cuts and corrections, as necessary, to optimize their analysis.
This compilation also serves as a fiducial sample --- to enable straightforward cross-comparison,
users should run their analysis on the full
sample with the value-added products fully implemented (\S~\ref{sec:data}), 
in addition to analyses incorporating alternative cuts, corrections, or
continuum normalizations.  The BOSS Collaboration has adopted this
strategy for our \lya\ forest BAO analysis.

The BOSS DR9 \lya\ Forest Sample is an unprecedented data set: 
it encompasses a co-moving volume of $\sim 20\;h^{-3}\,{\rm Gpc^3}$ 
and represents a dense sampling at $\sim 16$ quasar sightlines per square degree.
We hope that readers who have not previously worked with \lya\ forest data will take advantage of this
unique data set to make their own contribution to our understanding of the high-redshift universe.